\documentclass[iop,revtex4]{emulateapj}
\usepackage{lscape} 
\usepackage{natbib}

\usepackage{rotating}
\usepackage{amsmath}
\newcommand \kms{km~$\rm{s}^{-1}$}

\newcommand \msol{M$_{\odot}$}

\newfont{\rten}{cmr10} 
\def \arcdeg{\hbox{$^\circ$}}

\begin{document}

\title{Origin of 1I/'Oumuamua. I. An ejected protoplanetary disk object?}

\author{Amaya Moro-Mart\'{\i}n$^{1}$}

\altaffiltext{1}{Space Telescope Science Institute, 3700 San Martin Dr., Baltimore, MD 21218, USA; email: amaya@stsci.edu}

\begin{abstract}

1I/'Oumuamua is the first interstellar interloper to have been detected. Because planetesimal formation and ejection of predominantly icy objects are common by-products of the star and planet formation processes, in this study we address whether 1I/'Oumuamua could be representative of this background population of ejected objects. The purpose of the study of its origin is that it could provide information about the building blocks of planets in a size range that remains elusive to observations, helping to constrain planet formation models.  We compare the mass density of interstellar objects inferred from its detection to that expected from planetesimal disks under two scenarios: circumstellar disks around single stars and wide binaries, and circumbinary disks around tight binaries. Our study makes use of a detailed study of the PanSTARRS survey volume; takes into account that the contribution from each star to the population of interstellar planetesimals depends on stellar mass, binarity, and planet presence; and explores a wide range of possible size distributions for the ejected planetesimals, based on solar system models and observations of its small-body population. We find that 1I/'Oumuamua is unlikely to be representative of a population of isotropically distributed objects, favoring the scenario that it originated from the planetesimal disk of a young nearby star whose remnants are highly anisotropic. Finally, we compare the fluxes of meteorites and micrometeorites observed on Earth to those inferred from this population of interstellar objects, concluding that it is unlikely that one of these objects is already part of the collected meteorite samples.
\end{abstract}


\section{INTRODUCTION}
\label{Introduction}

The presence of protoplanetary and debris disks indicates that planetesimal formation is a common by-product of the star formation process (Moro-Mart{\'{\i}}n \citeyear{2013pss3.book..431M} and references therein).  The discovery of thousands of extra-solar planetary systems is evidence that, in some cases, this has led to the formation of planets in a wide range of planetary architectures (Winn \& Fabrycky \citeyear{2015ARA&A..53..409W}). 

Observations and models of the solar system, addressing its formation and dynamical evolution, indicate that a significant fraction of the planetesimals that initially formed during the planet formation process were eventually ejected into interstellar space by gravitational perturbation with the planets. This happened during different stages of the solar system's evolution. It is thought that before 10 Myr after the Sun was formed, while the Sun was still embedded in its maternal cluster, and before the gas in the primordial protoplanetary disk dispersed, Jupiter and Saturn formed and scattered planetesimals in the Jupiter--Saturn region to large distances; a fraction of this material had their perihelion lifted beyond the influence of the giant planets due to external perturbations by the stars and the gas in the cluster, populating the Oort could; but most of the scattered material (75--85\%; Brasser et al. \citeyear{2006Icar..184...59B}) was ejected into interstellar space. At a latter stage, between 10 and 100 Myr,  it is thought that the gravitational perturbations from Jupiter and Saturn, and the mutual perturbations amongst the largest asteroids, depleted the asteroid belt by a factor of $\sim$100, leaving behind an asteroid belt about 10--20$\times$ more massive than today (O'Brien et al. \citeyear{2007Icar..191..434O}). And the latest episode of massive planetesimal clearing would have occurred at $\sim$ 700 Myr, when it is thought that the migration of the giant planets (due to the interaction of the planets with a massive trans-Neptunian planetesimal disk) triggered an instability that resulted in the rearrangement of the planets' orbits into their current configuration; in this process, secular resonances swept through the asteroid belt making their orbits unstable, scattering some asteroids into the inner solar system producing the Late Heavy Bombardment, while ejecting others out of the  system, resulting in an additional depletion factor of $\sim$10--20;  this planet rearrangement also resulted in the sudden outward migration of Neptune and its exterior mean motion resonances, that swept through the Kuiper belt heavily depleting it (Gomes et al. \citeyear{2005Natur.435..466G}; Morbidelli et al. \citeyear{2005Natur.435..462M}; Tsiganis et al. \citeyear{2005Natur.435..459T}). Overall, it is estimated that in the solar system only a very small fraction of negligible mass of the initial planetesimal disk was left behind. 

Even though the efficiency of planetesimal ejection is very sensitive to the planetary architecture and its dynamical history, dynamical models indicate that planetesimal clearing processes might be common under a wide range of architectures (Raymond et al. \citeyear{2018MNRAS.476.3031R} and references therein). This has led to the idea that the interstellar space must be filled with objects with a planetesimal disk origin, a population that would have a  predominantly icy composition because the majority of the ejected material would have formed outside the snowline in their parent systems. 

Some of these objects would enter the solar system in highly hyperbolic orbits due to the the velocity of the Sun with respect to the Local Standard of Rest (LSR; $v_{\rm LSR}$ = 16.5 \kms), making these interstellar interlopers clearly distinguishable from other solar system objects. However, up to the discovery of 1I/'Oumuamua, not a single object with these kinematic properties had been discovered, triggering studies to address if the absence of detections was consistent with our expectations from planetesimal formation models and observations (McGlynn \& Chapman \citeyear{1989ApJ...346L.105M}; Moro-Mart\'{\i}n et al. \citeyear{2009ApJ...704..733M}; Engelhardt et al. \citeyear{2017AJ....153..133E}). 

The detection of 1I/'Oumuamua by PanSTARRS (Williams \citeyear{Williams2017}), with a clearly hyperbolic orbit (eccentricity $e$ = 1.197, semi-major axis $a$ =  -1.290, perihelion $q$ = 0.254, and inclination $i$ = 122.6), and high pre-encounter velocity (26.22 \kms, with U, V, W = -11.325, -22.384, -7.629 \kms; Mamajek \citeyear{2017RNAAS...1a..21M}), had therefore been been anticipated for decades.  Its composition, initially identified as refractory because of the lack of cometary activity (Jewitt et al. \citeyear{2017ApJ...850L..36J}; Meech et al. \citeyear{2017Natur.552..378M}), but now established as icy because of the evidence of outgassing (the lack of activity likely due to the presence of a thin insulating mantle; Micheli et al. \citeyear{2018Natur.559..223M}), also agrees well with the expectation that the majority of the ejected planetesimals populating the interstellar medium would have formed beyond the snowline of their parent systems. 

1I/'Oumuamua's lightcurve (with 2--2.5 photometric variability) indicates that its shape is very elongated, with different authors estimating an axis ratio ranging from 3 to 10 ($>$ 6:1 in Jewitt et al. \citeyear{2017ApJ...850L..36J}, 5:3.1 in Banninster et al. \citeyear{2017ApJ...851L..38B}, 10:1 in Meech et al. \citeyear{2017Natur.552..378M}, $>$ 4.63 in Drahus et al. \citeyear{2018NatAs...2..407D}, from 3.5 to 10.3 in Bolin et al. \citeyear{2018ApJ...852L...2B}, $>$ 5:1 in Fraser et al. \citeyear{2018NatAs...2..383F}). For the radius, estimates range from  55 m (for albedo 0.1, Jewitt et al. \citeyear{2017ApJ...850L..36J}), 60 m (for albedo 0.04, Fraser et al. \citeyear{2018NatAs...2..383F}), 102 m (for albedo 0.04, Meech et al. \citeyear{2017Natur.552..378M}), and 130 m (for albedo 0.03, Bolin et al. \citeyear{2018ApJ...852L...2B}), the uncertainties arising from its unknown shape and albedo. Its tumbling state and composition are consistent with a bulk density of $\sim$ 1 g/cm$^3$ (Drahus et al. \citeyear{2018NatAs...2..407D}). 1I/'Oumuamua is a small object that was fortuitously detected in a magnitude-limited survey after its perihelion passage because it happened to pass very close to the Earth on its way out of the solar system (Jewitt et al. \citeyear{2017ApJ...850L..36J}).  
	
Different studies have calculated the number density of 1I/'Oumuamua-like objects, using a range of estimates for the detection volume, and inferred from these estimates what the contribution per star would need to be, discussing whether or not these estimates agree with expectations (Gaidos et al. \citeyear{2017RNAAS...1a..13G}; Laughlin \& Batygin \citeyear{2017RNAAS...1a..43L}; Trilling et al., \citeyear{2017ApJ...850L..38T}; Do et al. \citeyear{2018ApJ...855L..10D}; Feng \& Jones \citeyear{2018ApJ...852L..27F}; Rafikov \citeyear{2018ApJ...861...35R}; Raymond et al. \citeyear{2018MNRAS.476.3031R}; Portegies-Zwart et al. \citeyear{2018MNRAS.479L..17P}).

The study presented here addresses the same overall question. The goal is to shed light on the potential origin of 1I/'Oumuamua by comparing the total mass density of interstellar objects inferred from its detection, assuming the object is representative of a population that is isotropically distributed, to the total mass density expected from the ejection of planetesimal from circumstellar and circumbinary disks. Our approach differs from previous studies in that: (1) it makes use of a detailed study of the PanSTARRS survey volume; (2) it takes into account that the contribution from each star to the population of interstellar objects depends on the mass of the central mass, whether the star is in a single or a close binary system (where "singles" also refer to wide binaries), and whether it is a planet-host; and (3) it takes into account that the ejected planetesimals populating the interstellar medium have a size distribution, for which we have assumed a wide range of possible distributions based on solar system models and observations, approximated as broken power laws. This approach addresses some of the caveats of previous studies. 

Based on a single detection, it is difficult to predict when we will be able to observe the next interstellar interloper. But 1I/'Oumuamua, as a true "messenger from afar arriving first" (the meaning of its name in Hawaiian), has opened a new window to study the planet building blocks around other stars in a size range that is not accessible via remote observations of these systems (tracing only the two extremes of the size distribution: the dust component and the planets). Its study can therefore complement the information we have from the study of meteorites, asteroids, Kuiper belt objects, comets, and other minor bodies in our solar system, and from the study of protoplanetary and debris disks around other stars, setting constraints on planetesimal and planet formation models. 

\section{Expected Mass Density of Interstellar Objects Based on the Detection of 1I/'Oumuamua}

\subsection{Cumulative Number Density of Interstellar Objects}
\label{NumberDensity}
To estimate the total mass density of interstellar objects, we adopt the number density distribution derived by Do et al. (\citeyear{2018ApJ...855L..10D}). They make the assumption that the objects are isotropically distributed, and adopt for 1I/'Oumuamua an absolute magnitude of $H$ = 22.1, a nominal phase function with slope parameter $G$ = 0.15, and a velocity at infinity, $v_{\infty}$ = 26 \kms. For these values, they compute the minimum and maximum distance that PanSTARRS could have seen such an object, and assume that each observation covers 6 deg$^2$. They then calculate the total survey volume by taking into account the effect of gravitational focusing by the Sun, trailing losses due the the tangential velocity of the object with respect to the Earth, the degradation of the signal by the background noise, that the object can come from any approach direction, and assuming that the detection rate for objects as large as 1I/'Oumuamua is given by one detection in the 3.5 year survey time. Assuming that the cumulative number density of interstellar objects down to the detection size $R$ (taken as 1I/'Oumuamua's size) is the inverse of the survey volume, they estimate a cumulative number density of $N_{\rm r \geqslant R}$ =  0.21 au$^{-3}$ $\sim$ 2 $\cdot~10^{15}$ pc$^{-3}$. They note that this is an underestimate of at most 40\% because the objects have a cumulative size distribution that falls at larger sizes. On the other hand, they point out that the detection process is not 100\% efficient over the full 6 deg$^2$ and given these inefficiencies the detection volume could be 2/3--3/4 of the nominal value, so that the number density could be 4/3--3/2 of their inferred number density. 

For comparison, other authors estimate that the number density is 0.1 au$^{-3}$ = 8$\cdot10^{14}$ pc$^{-3}$ (Jewitt et al. \citeyear{2017ApJ...850L..36J}, Fraser et al. \citeyear{2018NatAs...2..383F}), 0.012--0.087 au$^{-3}$ = 1--7$\cdot10^{14}$ pc$^{-3}$ (Portegies-Zwart et al. \citeyear{2018MNRAS.479L..17P}), 0.012 au$^{-3}$ = 1$\cdot10^{14}$ pc$^{-3}$ (Gaidos et al. \citeyear{2017RNAAS...1a..13G}), and $<$ 0.006 au$^{-3}$ = 4.8$\cdot10^{13}$ pc$^{-3}$ (lower limit from Feng \& Jones \citeyear{2018ApJ...852L..27F}). These estimates assume a range of survey times (e.g. 1--2 years Jewitt et al. \citeyear{2017ApJ...850L..36J}; 5 years Portegeis-Zwart et al. \citeyear{2018MNRAS.479L..17P}; 7 years Gaidos et al. \citeyear{2017RNAAS...1a..13G}; 20 years Feng \& Jones \citeyear{2018ApJ...852L..27F}) and also a small range of dark albedos and absolute magnitude that result in a range of average object radius (55 m Jewitt et al. \citeyear{2017ApJ...850L..36J}; 60 m Fraser et al. \citeyear{2018NatAs...2..383F}; 100 m Portegeis-Zwart et al. \citeyear{2018MNRAS.479L..17P}; 115 m Gaidos et al. \citeyear{2017RNAAS...1a..13G}; 50 m Feng \& Jones \citeyear{2018ApJ...852L..27F}). We adopt the cumulative number density estimate from Do et al. (\citeyear{2018ApJ...855L..10D}) because of their careful calculation of the PanSTARRS detection volume. 

Using this number density, we estimate below the expected mass density of interstellar objects but, as opposed to the studies mentioned above, we explore a wide range of possible size distributions. 

We note that the value of $N_{\rm r \geqslant R}$ that we are adopting from Do et al. (\citeyear{2018ApJ...855L..10D}) assumes mono-sized objects. Ideally, in the calculations described below, one would want to recalculate this value for the wide range of size distributions considered. However, given that from Do et al. (\citeyear{2018ApJ...855L..10D}) we are adopting the order of magnitude estimate of $N_{\rm r \geqslant R}$ $\sim$ 2 $\cdot~10^{15}$ pc$^{-3}$, and that that in the parameter space exploration described in Section \ref{ResultingMassDensity}, we are studying the effect of adopting a wide range of other possible cumulative number densities (based on the values mentioned above, inferred by other authors), we will adopt the Do et al. (\citeyear{2018ApJ...855L..10D})  value as such but warn the reader that, in doing so, an approximation is being made.

\subsection{Size Distributions Considered for the Interstellar Object Population}
\label{SizeDistribution}
The only small-body population that can be studied in any detail is that of the solar system. Assuming initially that the source of the interstellar objects are planetesimal disks similar to that of the early solar system, and using models and observations of its small-body population, together with theoretical models of accretion processes, we explore the following range of possible size distributions, where $r$ is the object radius. 

\subsubsection{Power law Size Distribution with Two Slopes}
\label{TwoSlopes}

\begin{equation}
\begin{split}
n(r) \propto r^{-q_1}~{\rm if}~r_{min}<r<r_b\\
n(r) \propto r^{-q_2}~{\rm if}~r_b<r<r_{max}\\
q_1 = {\rm 2.0,~2.5,~3.0,~3.5}\\
q_2 = {\rm 3,~3.5,~4,~4.5,~5}\\
r_b = {\rm 3~km, 30~km,~90~km}\\
r_{max} \approx~{\rm1000~km~and}~r_{min} \approx~{\rm1}~\mu{\rm m}\\
\end{split}
\label{2s-sizedist}
\end{equation}

This selection is similar to that adopted in Moro-Mart\'{\i}n et al. (\citeyear{2009ApJ...704..733M}) and is based on solar system observations and on accretion and collisional models. 

From the current size distribution of the asteroid belt, and taking into account observational constraints and collisional evolution models, Bottke et al. (2005) estimated that the size distribution in the ``primordial'' asteroid followed a broken power law of  
$n(r) \propto r^{-q_1}~{\rm if}~r<r_b$ and 
$n(r) \propto r^{-q_2}~{\rm if}~r>r_b$, 
where {\it r} is the planetesimal radius, {\it r$_b$} $\approx$ 50 km, {\it q$_1$} $\approx$ 1.2 and {\it q$_2$} $\approx$ 4.5. 
This distribution would have been established early on as a result of a period of collisional activity before 
Jupiter formed (few Myr), and a period of collisional activity  triggered by the planetary embryos (10--100 Myr).

For the Kuiper belt (KB), the Hubble Space Telescope survey by Bernstein et al. (\citeyear{2004AJ....128.1364B}) yielded {\it q$_1$} = 2.9 and {\it q$_2$} $>$ 5.85 for the classical KB, and {\it q$_1$} $<$ 2.8 and {\it q$_2$} = 4.3 for the excited KB, with {\it r$_b$} $\leq$ 50 km in both cases; a pencil-beam search by Fuentes, George \& Holman (\citeyear{2009ApJ...696...91F}) found {\it r$_b$} $\approx$ 45$\pm$15($p$/0.04)$^{-0.5}$ km (where $p$ is the albedo); and a Subaru survey by Fraser and Kavelaars (\citeyear{2009AJ....137...72F}), sensitive to KBOs with $r >$ 10 km, found {\it q$_1$} = 1.9, {\it q$_2$} = 4.8, and {\it r$_b$} $\leq$ 25--47 km (assuming a 6\% albedo). Other KBO surveys reviewed by Kenyon et al. (\citeyear{2008ssbn.book..293K}) yielded $q_2$ $\approx$ 3.5--4 and {\it r$_{max}$} $\sim$ 300--500 km for the cold classical KB  ({\it a} = 42--48 au, {\it perihelion} $>$37 au, {\it i} $\lesssim$ 4$^\circ$); $q_2$ $\approx$ 3 and {\it r$_{max}$} $\sim$ 1000 km for the hot classical KB ({\it i} $\gtrsim$ 10$^\circ$); and  $q_2$ $\approx$ 3 and {\it r$_{max}$} $\sim$ 1000 km for the resonant population; in all cases,  the transitional radius is {\it r$_b$  $\approx$} 20--40 km (for albedo $\sim$ 0.04--0.07). 

Elliptic comets follow a power law of index $q \approx$ 2.9 for $r >$ 1.6 km or $q \approx$ 2.6, when including cometary near-Earth objects (thought to be extinct elliptic comets; Lamy et al. \citeyear{2004come.book..223L}). Because their size distribution has heavily evolved due to collisions and perihelion passages, it might be more representative of systems that have experienced a high degree of collisional activity. 

Coagulation models that take into account the collisional evolution due to self-stirring find that the differential size distribution expected for KBOs  follows a broken power law with
{\it n(r) $\propto$ r$^{-q_1}$} if {\it r $\leq$ r$_1$},  
{\it n(r) $\propto$ \rm constant} if {\it r$_1$ $\leq$ r $<$ r$_0$}, and
{\it n(r) $\propto$ r$^{-q_2}$} if {\it r $\geq$ r$_0$}, 
where {\it r} is the planetesimal radius and $q_1$ $\approx$ 3.5 (resulting from the collisional cascade);  for a fragmentation parameter, {\it Q$_b$} $\gtrsim$ 10$^5$ erg g$^{-1}$,  $q_2$ $\approx$ 2.7--3.3, and  {\it r$_0$  $\approx$ r$_1$ $\approx$} 1 km;  while for {\it Q$_b$} $\lesssim$ 10$^3$ erg g$^{-1}$, $q_2$ $\approx$ 3.5--4, {\it r$_1$  $\approx$} 0.1 km, and {\it r$_0$  $\approx$} 10--20 km (see review in Kenyon et al. 2008). These authors argue that the comparison between the observational and the modeling results indicates that self-stirring alone cannot account for the observed size distribution in the KB and that dynamical perturbations have played a major role, suggesting that the size distribution in the KB was frozen after a short-lived event of dynamical excitation likely produced by planet migration (otherwise, it would have evolved to a shallower distribution). Fraser and Kavelaars (\citeyear{2009AJ....137...72F}) argued that the large {\it r$_b$} observed implies increased collisional evolution,  while the large {\it q$_2$} suggests there was an early end to the accretion stage. 

None of these models can simultaneously reproduce the observed slopes for the large and the small objects and the break radius, probably because they do not take into account the full dynamical history of the system. Other planetary systems, potential sources of interstellar objects, will likely have experienced a wide range of dynamical and collisional histories, and the size distribution of their ejected bodies will therefore depend significantly on the degree of dynamical/collisional evolution at the time of the ejection. To take this into account, we consider a wide range of possible size distributions, based on the observations and the models described above, encompassed by the range of parameters in Equation \eqref{2s-sizedist}. 

\subsubsection{Power law Size Distribution with Five Slopes}
\label{FiveSlopes}
\begin{equation}
\begin{split}
n(r) \propto r^{-q_1}~{\rm if}~r_{min}<r<r_{b1}\\
n(r) \propto r^{-q_2}~{\rm if}~r_{b1}<r<r_{b2}\\
n(r) \propto r^{-q_3}~{\rm if}~r_{b2}<r<r_{b3}\\
n(r) \propto r^{-q_4}~{\rm if}~r_{b3}<r<r_{b4}\\
n(r) \propto r^{-q_5}~{\rm if}~r_{b4}<r<r_{max}\\
r_{max} \approx~{\rm1000~km~and}~r_{min} \approx~{\rm1}~\mu{\rm m}
\end{split}
\label{5s-sizedist}
\end{equation}

In both cases, ${\it r}_{\rm1}$ = 0.1 km, ${\it r}_{\rm2}$ = 2 km, ${\it r}_{\rm3}$ = 10 km, and ${\it r}_{\rm4}$ = 30 km, with slopes:
\begin{itemize}
\item At 4.5 Gyr, ${\it q}_{\rm1}$ = 3.7, ${\it q}_{\rm2}$ = 2.5, ${\it q}_{\rm3}$ = 5.8, ${\it q}_{\rm4}$ = 2.0, ${\it q}_{\rm5}$ = 4.01. 
\item At 100 Myr, ${\it q}_{\rm1}$ = 3.7, ${\it q}_{\rm2}$ = 2.6, ${\it q}_{\rm3}$ = 6.8, ${\it q}_{\rm4}$ = 3.2, ${\it q}_{\rm5}$ = 3.8. 
\end{itemize}

This five-slope power law is based on Schlichting et al. (\citeyear{2013AJ....146...36S}), who studied the size distribution of the small-body population of the Kuiper belt. They find that the current size distribution can be explained by coagulation models starting with an initial planetesimal population of $\sim$ 1 km radius objects (where the best match is for an initial population containing about equal mass per logarithmic mass bin in bodies ranging from 0.4 km to 4 km), and subsequent collisional evolution. The size distribution beyond $\sim$ 30 km would be primordial, i.e. a relic from the the accretion history, and is well matched by coagulation models of runaway growth. While the size distribution below 30 km would have been modified by collisional evolution, with a slope that changes with time and object radius, approaching the value expected for an equilibrium collisional cascade of material-strength dominated bodies for objects $<$ 0.1 km (about the size of 1I/'Oumuamua).
Compared to a single power law size distribution, there is a strong excess of bodies at $r \sim$ 2 km, caused by the planetesimal size distribution left over from the runaway growth phase, that remains even after 4.5 Gyr of collisional evolution, and a strong deficit of bodies with $r \sim$ 10 km, a depletion caused by collisions with the excess population of km size bodies. 

In our study, to take into account the possible different degrees of collisional evolution, we will adopt the size distribution resulting from the models in Schlichting et al. (\citeyear{2013AJ....146...36S})  at two different times: 100 Myr and 4.5 Gyr [the corresponding parameters are listed in Equation \eqref{5s-sizedist}]. 

A caveat of using this power law is that it reflects that, in the solar system, the formation and migration of the outer planets and resulting excitation of the velocity dispersion of the growing planetesimals halted planet formation in the KB region, while in other systems this formation history may have been different, changing the resulting size distribution. On the other hand, it makes the KB an interesting study case precisely because it is a remnant of the primordial system (Schlichting et al. \citeyear{2013AJ....146...36S}), and the the source of the 1I/'Oumuamua-like interstellar objects might be young systems. 

Another caveat is that the size distribution depends on the strength law and the catastrophic destruction threshold ($Q_D^*$, the specific energy needed to create a spectrum of targets where the largest one has half the mass of the initial target), so we are assuming that the small-body population in other systems are similar to those in the solar system, which might not necessarily be the case, in particular given the remarkably elongated shape of 1I/'Oumuamua. However, Schlichting et al. (\citeyear{2013AJ....146...36S}) point out that the main signatures (the overall excess of $\sim$ 2 km size bodies and the deficit of $\sim$ 10 km size bodies) are independent of these parameters and the initial size distribution (which remains as one of the major open questions), making this size distribution worth exploring. 

\subsection{Bulk Density}
\label{BulkDensity}

To estimate the total mass density of interstellar objects, using the number density derived in $\S$ \ref{NumberDensity} and  the range of size distributions discussed in $\S$ \ref{SizeDistribution}, we need to make an assumption for the object's bulk density. Due to the lack of cometary activity, some authors had argued that 1I/'Oumuamua might be of refractory nature and asteroidal origin (Trilling et al., \citeyear{2017ApJ...850L..38T}; Jackson et al. \citeyear{2018MNRAS.477L..85J}; Rafikov \citeyear{2018ApJ...861...35R}; Raymond et al. \citeyear{2018MNRAS.476.3031R}). However, its surface physical properties, that resemble that of a comet (Jewitt et al. \citeyear{2017ApJ...850L..36J}; Fitzsimmons et al. \citeyear{2018NatAs...2..133F}), its tumbling state that is consistent with a low bulk density (Drahus et al. \citeyear{2018NatAs...2..407D}), and its recently found nongravitational acceleration, consistent with outgassing  (Micheli et al. \citeyear{2018Natur.559..223M}), all indicate that the object is icy and of cometary origin.  Therefore, we will assume a bulk density of 1.0 g/cm$^3$, following Drahus et al. (\citeyear{2018NatAs...2..407D}). 

\subsection{Object Shape and Radius}
\label{ObjectRadius}
1I/'Oumuamua's lightcurve indicates that its shape is elongated, with different authors estimating an axis ratio ranging from 3 to 10 ($>$ 6:1 in Jewitt et al. \citeyear{2017ApJ...850L..36J}, 5:3.1 in Banninster et al. \citeyear{2017ApJ...851L..38B}, 10:1 in Meech et al. \citeyear{2017Natur.552..378M}, $>$ 4.63 in Drahus et al. \citeyear{2018NatAs...2..407D}, from 3.5 to 10.3 in Bolin et al. \citeyear{2018ApJ...852L...2B}, $>$ 5:1 in Fraser et al. \citeyear{2018NatAs...2..383F}). For the radius, estimates range from  55 m (for albedo 0.1, Jewitt et al. \citeyear{2017ApJ...850L..36J}), 60 m (for albedo 0.04, Fraser et al. \citeyear{2018NatAs...2..383F}), 102 m (for albedo 0.04, Meech et al. \citeyear{2017Natur.552..378M}), and 130 m (for albedo 0.03, Bolin et al. \citeyear{2018ApJ...852L...2B}). In our study, we adopt initially an intermediate effective ratio of 80 m, corresponding to an albedo of 0.037 and axis ratio $>$ 4.63, estimated by Drahus et al. (\citeyear{2018NatAs...2..407D}), but also explore other radii because of uncertainties in the albedo and the shape of the object. For example, Fraser et al. (\citeyear{2018NatAs...2..383F}) estimate an average radio of 60 m for albedo 0.04, but point out that the albedo could be as high as 0.08, with the corresponding decrease in object size.

\subsection{Resulting Mass Density}
\label{MassDensity}

\subsubsection{From the Power Law Size Distribution with Two Slopes}
\label{MassDensityTwoSlopes}
Using Equation \eqref{2s-mtotal} in Appendix A.1 (derived for a two-slope size distribution), where $R$ and $r_{\rm i}$ are in cm, we can estimate the expected mass density of interstellar objects, $m_{\rm total}$ (g pc$^{-3}$), using the cumulative number density discussed in $\S$ \ref{NumberDensity} ($N_{\rm r \geqslant R}$ = 2 $\cdot~10^{15}$ pc$^{-3}$),  the bulk density discussed in $\S$ \ref{BulkDensity} ($\rho$ = 1 g cm$^{-3}$), the object radius discussed in $\S$ \ref{ObjectRadius} ($R$ = 80 m), and by exploring the range of parameters for the two-slope power law size distribution described in Equation \ref{2s-sizedist}. 

\begin{figure}
\begin{center}
\includegraphics[width=6cm]{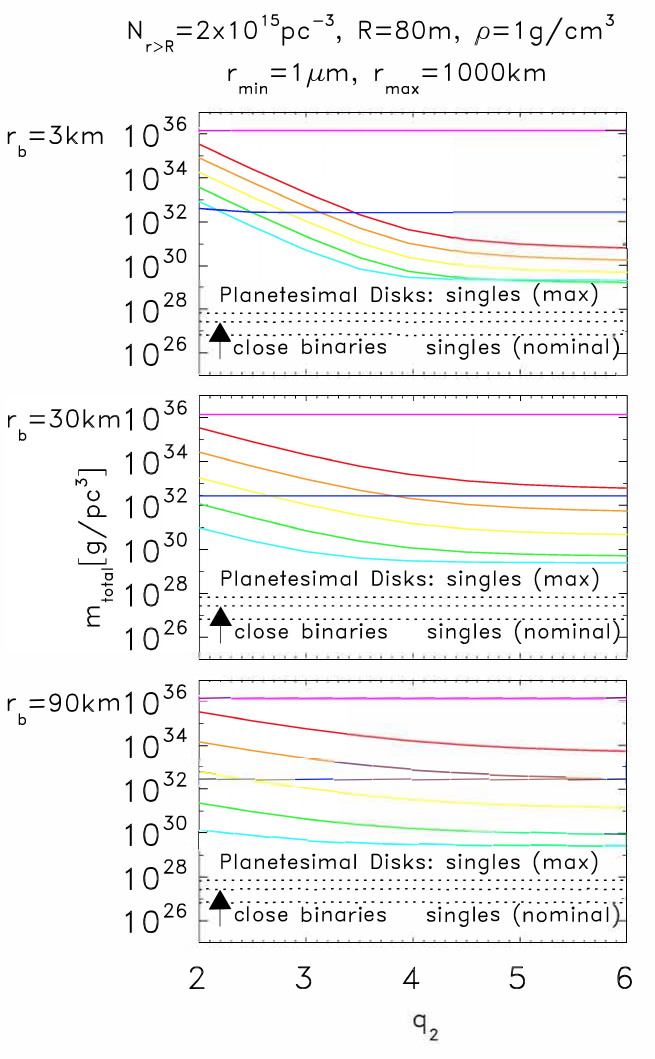}
\end{center}
\caption{Expected mass density of interstellar objects, $m_{\rm total}$ (g pc$^{-3}$), calculated from Equation \eqref{2s-mtotal}, with $N_{\rm r \geqslant R}$ = 2 $\cdot~10^{15}$ pc$^{-3}$ (Do et al. \citeyear{2018ApJ...855L..10D}), $R$ = 80 m (corresponding to an albedo of 0.037; Drahus et al. \citeyear{2018NatAs...2..407D}), $\rho$ = 1 g cm$^{-3}$ (Drahus et al. \citeyear{2018NatAs...2..407D}), and for the range of power law slopes ($q_{\rm 1}$ and $q_{\rm 2}$) and break radius ($r_{\rm b}$) discussed in Equation \ref{2s-sizedist}. The different colors correspond to different values of $q_{\rm 1}$, with $q_{\rm 1}$ = 2 (red), 2.5 (orange), 3 (yellow), 3.5 (green), 3.95 (light blue; used instead of 4 to avoid dividing by zero), 4.5 (dark blue), and 5 (pink). The dotted lines correspond to the mass density of interstellar objects expected to arise from planetesimal disks around stars in single and wide binary systems [discussed in $\S$ \ref{ResultingMassDensity}, adding Equations  \eqref{MassDensitySinglesMK2} and \eqref{MassDensitySinglesK2A}],  and from tight binary systems [discussed in $\S$ \ref{ResultingMassDensity} and given by Equation  \eqref{MassDensityBinaries}]. The top line, labeled  {\it singles (max)}, corresponds to the maximum mass density of interstellar material  expected to originate from single and wide binary systems if we were to assume that all stars, independently on whether or not they host a planet, contribute [i.e. if we adopt $f_{\rm pl}$ = 1 in Equations \eqref{MassDensitySinglesMK2} and \eqref{MassDensitySinglesK2A}]. The bottom line, labeled {\it singles (nominal)}, is the contribution expected from single stars and wide binaries in the nominal case under which only stars with giant planets contribute [or Neptune-size planets in the case of low-mass stars; i.e. if we adopt $f_{\rm pl}$ = 0.03 in Equation \eqref{MassDensitySinglesMK2} and $f_{\rm pl}$ = 0.2 in Equation \eqref{MassDensitySinglesK2A}]. In both cases, for the top and the bottom lines, most of the material ejected would be icy. The middle line, labeled {\it close binaries}, corresponds to the expected contribution from tight binary systems. In this case, it is expected that about 36\% of the ejected material would be icy (Jackson et al. \citeyear{2018MNRAS.477L..85J}), so when considering only icy material (more relevant because of the icy nature of 1I/'Oumuamua), this estimate would be slightly larger than the {\it singles (nominal)} case.
}
\label{2s-nom}
\end{figure}

\begin{figure}
\begin{center}
\includegraphics[width=6cm]{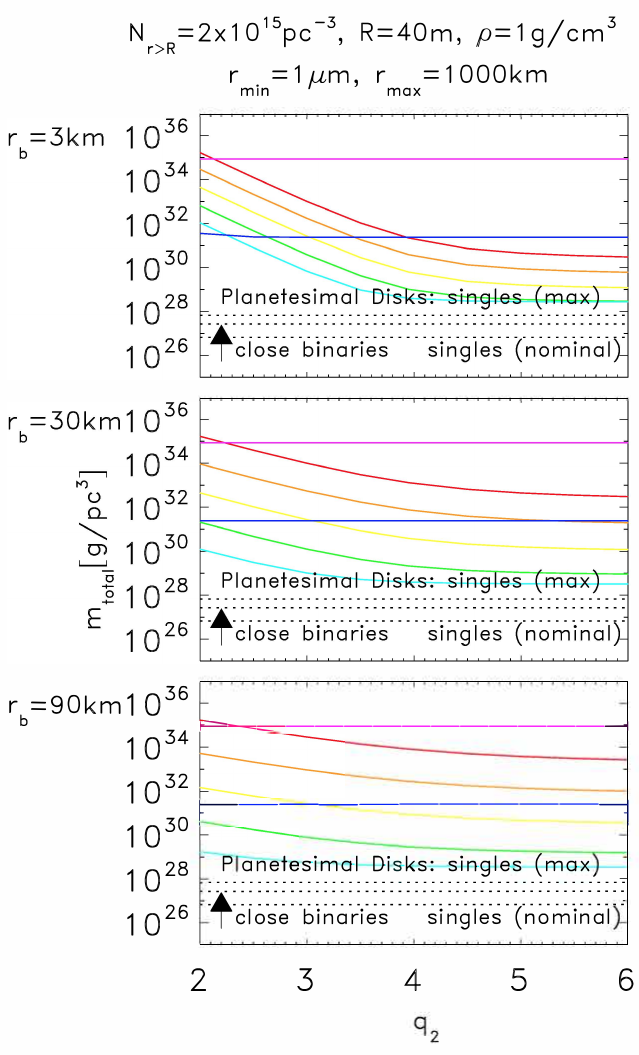}
\end{center}
\caption{Same as Figure \ref{2s-nom} but assuming a smaller effective radius of $R$ = 40 m (if the object had an albedo higher than the 0.037 assumed, as suggested by other authors). 
}
\label{2s-R40}
\end{figure}

\begin{figure}
\begin{center}
\includegraphics[width=6cm]{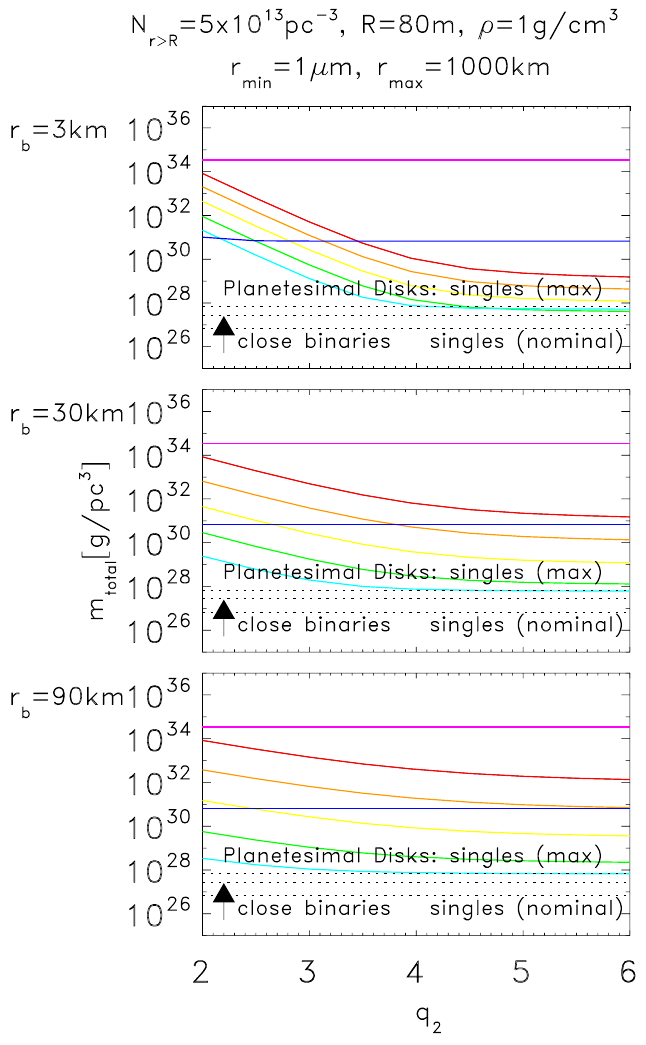}
\end{center}
\caption{Same as Figure \ref{2s-nom} but assuming the minimum value for the cumulative number density derived by Feng \& Jones (\citeyear{2018ApJ...852L..27F}), $N_{\rm r \geqslant R}$ $\sim$ 5 $\cdot10^{13}$ pc$^{-3}$. 
}
\label{2s-N5E13}
\end{figure}

\begin{figure}
\begin{center}
\includegraphics[width=6cm]{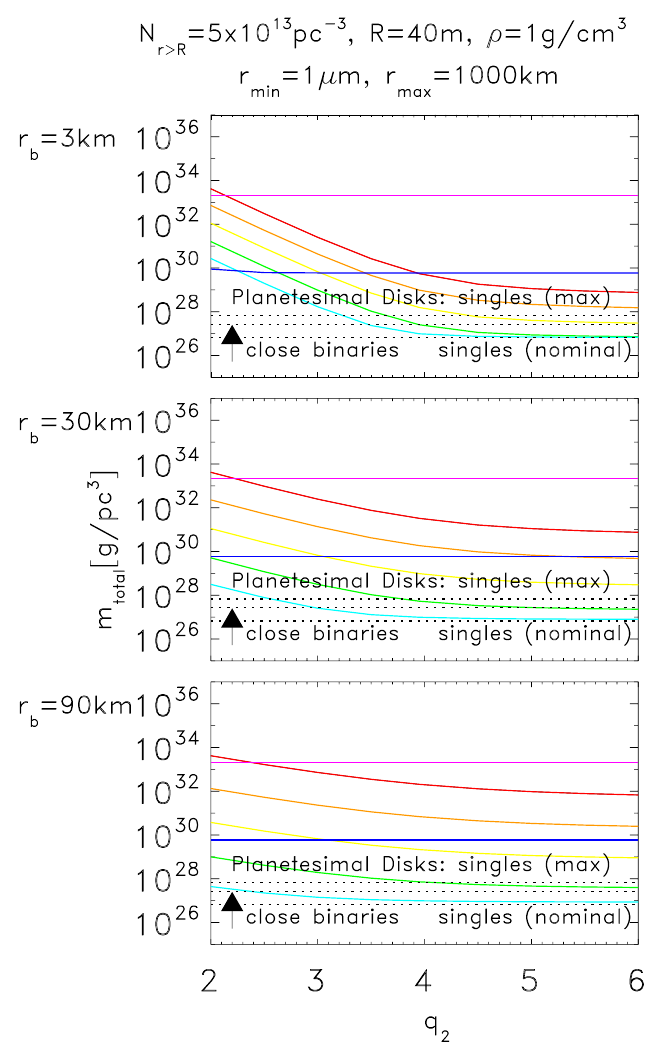}
\end{center}
\caption{Same as Figure \ref{2s-nom} but with $N_{\rm r \geqslant R}$ $\sim$ 5 $\cdot10^{13}$ pc$^{-3}$ and $R$ = 40 m. 
}
\label{2s-R40N5E13}
\end{figure}

Figures \ref{2s-nom}, \ref{2s-R40}, \ref{2s-N5E13} and \ref{2s-R40N5E13} show the results for the above nominal values, in addition to those that would arise if adopting lower values of the cumulative number density ($N_{\rm r \geqslant R}$) and object radius ($R$), in agreement with other authors. 
The dotted lines in the Figures correspond to the mass density of interstellar objects expected from a planetesimal disk origin (discussed below in $\S$ \ref{ExpectedMassDensity}). The results are discussed in $\S$ \ref{Comparison}.

\subsubsection{From the Power Law Size Distribution with Five Slopes}
\label{MassDensityFiveSlopes}

Similarly to $\S$ \ref{MassDensityTwoSlopes}, we now estimate the expected mass density of interstellar objects for the case where the size distribution is the one derived by  Schlichting et al. (\citeyear{2013AJ....146...36S}) for the solar system at 100 Myr (Figure \ref{5s-100M}) and at 4.5 Myr (Figure \ref{5s-4G}), using Equation \eqref{5s-mtot} in Appendix A.2. In this case, the slopes and boundary radius of the adopted size distribution remain fixed at the values given by Equation \eqref{5s-sizedist} (two different set of values for the 100 Myr and 4.5 Gyr cases, respectively). The axes of Figures \ref{5s-100M} and \ref{5s-4G} explore different values of the cumulative number density ($N_{\rm r \geqslant R}$) and the object radius ($R$) in logarithmic scale, with the colors corresponding to the calculated value of the mass density of interstellar objects ($m_{\rm total}$) for each pair of values. The mass density corresponding to 1I/'Oumuamua's nominal values ($N_{\rm r \geqslant R}$ = 2 $\cdot~10^{15}$ pc$^{-3}$,  $R$ = 80 m) is shown as a cross. For all cases, $\rho$ = 1 g cm$^{-3}$. The dotted lines are the same as in Figures 1--4. The results are discussed in $\S$ \ref{Comparison}.

\begin{figure}
\begin{center}
\includegraphics[width=6cm]{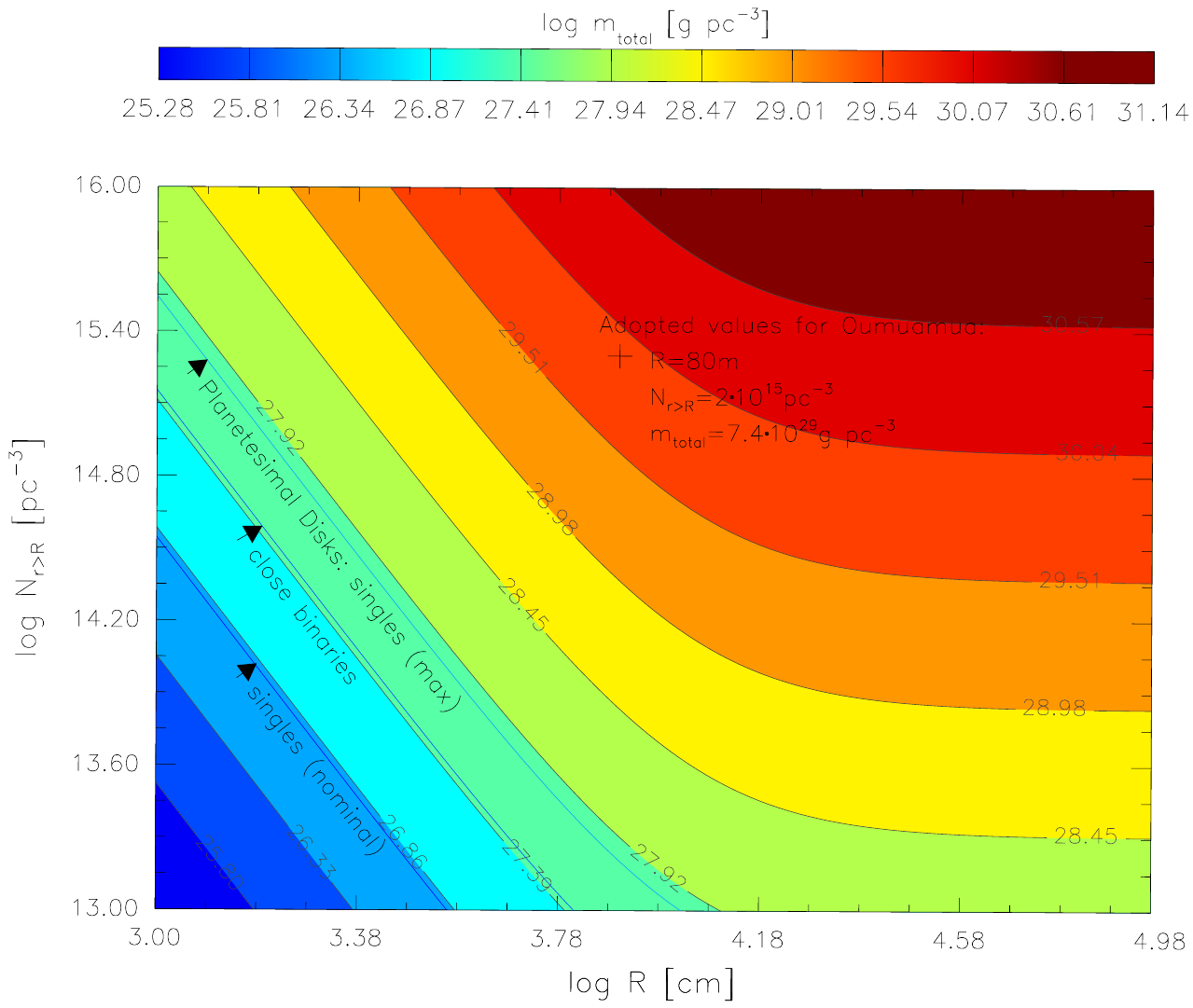}
\end{center}
\caption{Expected mass density of interstellar objects calculated from Equation \eqref{5s-mtot} in Appendix A.2, corresponding to the size distribution derived by  Schlichting et al. (\citeyear{2013AJ....146...36S}) for the solar system at 100 Myr (described by Equation \ref{5s-sizedist}). The $x$-axis corresponds to the cumulative number density in pc$^{-3}$ and the y-axis to the object radius in cm (both in log scale). The colors correspond to the calculated value of the mass density of interstellar objects ($m_{\rm total}$, in g pc$^{-3}$) for a given pair of values. The cross denotes the mass density corresponding to 1I/'Oumuamua's nominal values of $N_{\rm r \geqslant R}$ = 2 $\cdot~10^{15}$ pc$^{-3}$ and  $R$ = 80 m. The adopted bulk density is $\rho$ = 1 g cm$^{-3}$. The dotted lines correspond to the mass density of interstellar objects expected from a planetesimal disk origin (see caption of Figure \ref{2s-nom}).
}
\label{5s-100M}
\end{figure}

\begin{figure}
\begin{center}
\includegraphics[width=6cm]{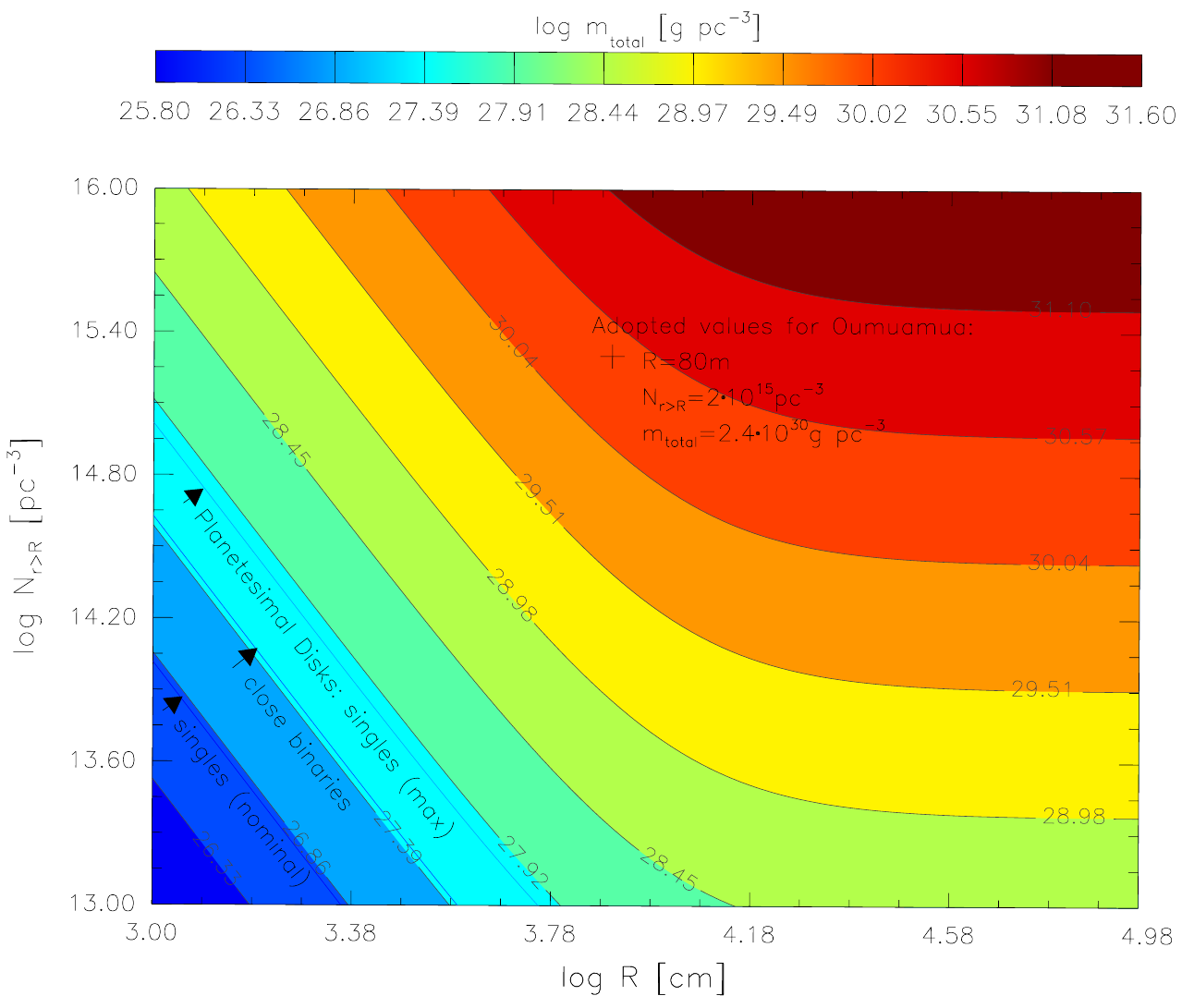}
\end{center}
\caption{Same as Figure \ref{5s-100M} but adopting the size distribution from Schlichting et al. (\citeyear{2013AJ....146...36S}) that corresponds to the solar system at 4.5 Gyr.
}
\label{5s-4G}
\end{figure}

\section{Expected Mass Density of Interstellar Objects with a Planetesimal Disk Origin}
\label{ExpectedMassDensity}

The hypothesis that we are testing in this study is whether the source of the population of interstellar objects could be the circumstellar and circumbinary planetesimal disks that form as part of the planetesimal/planet formation process, assuming that 1I/'Oumuamua is representative of a background population of interstellar objects that are isotropically distributed. To test this hypothesis, we now estimate the mass density of interstellar objects that would be expected to have a planetesimal disk origin, and compare that value to the one derived above from the detection of 1I/'Oumuamua (discussed in $\S$ \ref{MassDensity}). 

To do this, we follow the procedure in Moro-Mart{\'{\i}}n et al. (\citeyear{2009ApJ...704..733M}) with some modifications. This study differs from the approach taken by other authors 
(Gaidos et al. \citeyear{2017RNAAS...1a..13G}; Laughlin \& Batygin \citeyear{2017RNAAS...1a..43L}; Trilling et al., \citeyear{2017ApJ...850L..38T}; Do et al. \citeyear{2018ApJ...855L..10D}; Feng \& Jones \citeyear{2018ApJ...852L..27F}; Rafikov \citeyear{2018ApJ...861...35R}; Raymond et al. \citeyear{2018MNRAS.476.3031R}; Portegies-Zwart et al. \citeyear{2018MNRAS.479L..17P}) because, in addition to using the detailed study by Do et al. (\citeyear{2018ApJ...855L..10D}) of the PanSTARRS survey volume to estimate the cumulative number density of objects larger or equal than Oumuamua, it takes into account: (1) that the contribution from each star to the population of interstellar planetesimals depends on the mass of the central mass, whether the star is in a single, wide binary or tight binary system, and whether it is a planet-host; (2) a wide range of possible size distributions for the ejected planetesimals, based on solar system models and observations of its small-body population (as discussed in $\S$\ref{SizeDistribution}).  

We now calculate separately the contribution to the total mass density of interstellar solids from single stars and wide binary systems, and from tight binary systems. 

\begin{itemize}
\item Contribution from single stars and wide binary systems:\\
\begin{equation}
\begin{split}
m_{\rm total}^{\rm singles}=\\
\int \xi({\it M_{*}})10^{-4}{\it M_{*}}\left({{\it f_{\rm solids}} \over 10^{-4}}\right)\left({{\it f_{\rm eject}} \over 1}\right)(1-{\it f_{\rm bin}})\it{f_{\rm pl}}{\it dM_{*}},                
\end{split}
\label{mtotalSingles}
\end{equation}
\item Contribution from close binary systems:\\
\begin{equation}
\begin{split}
m_{\rm total}^{\rm binaries}= \\
\int \xi({\it M_{\rm sys}})10^{-4}{\it M_{\rm sys}}\left({{\it f_{\rm disk}} \over 0.1}\right)\left({{\it f_{\rm drift}} \over 0.1}\right)\left({{\it f_{\rm eject}} \over 1}\right){\it f_{\rm bin}}{\it dM_{\rm sys}},               
\end{split}
\label{mtotalBinaries}
\end{equation}
\end{itemize}
where the different terms are described in the following sections. 

\subsection{Number Density of Stars}
\label{NumberDensityStars} 
\subsubsection{Singles and Wide Binaries: $\xi(M_*)(1-f_{\rm bin})dM_*$}
Following Kroupa et al. (\citeyear{1993MNRAS.262..545K}), we assume that the number density of  stars per pc$^3$, out to $\sim$ 
130 pc from the Sun, within the mid-plane of the galaxy, and with stellar masses between {\it M$_{*}$} and {\it M$_{*}$+dM$_{*}$} (in units of M$_{\odot}$), is given by 
\begin{equation}
\begin{split}
n(M_*) = \xi(M_*)dM_*~~~~{\rm with,}~~~~~~~~~~~~\\ 
\xi(M_*) = 0.035 M_*^{-1.3}~~~~{\rm if~0.08} \le M_* < 0.5\\
\xi(M_*) = 0.019 M_*^{-2.2}~~~~~{\rm if~0.5} \le M_* < 1.0\\
\xi(M_*) = 0.019 M_*^{-2.7}~~~~{\rm if~1.0} \le M_* < 100.  
\end{split}
\label{eq_stellar_density}
\end{equation}
Equation \eqref{eq_stellar_density} corresponds to an initial mass function derived based on the present-day mass function, assuming no significant stellar evolution for low-mass stars, and using Scalo's (\citeyear{1986FCPh...11....1S}) initial mass function for higher masses (Kroupa et al. \citeyear{1993MNRAS.262..545K}). We use the initial mass function instead of the present-day mass function because we are interested in the contribution that the young planetesimal disks could have made to the population of interstellar planetesimals.
 
Following Jackson et al. (\citeyear{2018MNRAS.477L..85J}), we adopt a binary frequency, $f_{\rm bin}$ = 0.26, i.e. 26\% of stars are in tight binaries, while the remaining 74\% are in single or wide binary systems, where the boundary between wide and tight binaries is chosen at the point at which the outermost stable orbit around the more massive stars is $>$ 10 au. 

\subsubsection{Tight Binaries: $\xi(M_{\rm sys})f_{\rm bin}dM_{\rm sys}$}
We assume, following Jackson et al. (\citeyear{2018MNRAS.477L..85J}),  that the combined mass of the tight binary systems ($M_{\rm sys}$) is drawn from the same  population as the singles [described in Equation \eqref{eq_stellar_density}]. 

\subsection{Total Mass Available to Form Solids Per Star{\rm disk}}
\label{solids}
\subsubsection{Singles and wide binaries: $10^{-4}{\it M_{*}}(f_{\rm solids}/10^{-4})$}
For single stars and wide binaries, we will assume that the total amount of solid material (primordial dust) that surrounds each star and m{\rm disk}ight be available to accrete into larger bodies can be approximated as 10$^{-4}M_*$. This is based on submillimeter surveys of Class II sources in star-forming regions, which lead to similar disk mass  distributions, for example $<$M$_{\rm disk}> \sim$ 0.005 \msol~in Taurus ($<$age$>\sim$1 Myr) and $<$M$_{\rm disk}/M_*>\sim$ 0.01 in $\rho$-Oph$~(<$age$>\sim$0.7 Myr), where $M_{\rm disk}$ includes both gas and dust (Andrews \& Williams \citeyear{2007ApJ...671.1800A}). This is comparable to the minimum-mass solar nebula of $\sim$0.015 M$_{\odot}$ (the total disk mass  required to account for the condensed material in the solar system planets). Assuming a gas-to-dust ratio of 100:1 leads to a total of $\sim$10$^{-4}M_*$$(f_{\rm solids}/10^{-4})$ of solids per star, where we assume $f_{\rm solids}$ = 10$^{-4}$.

\subsubsection{Tight Binaries: $10^{-4}{\it M_{sys}}(f_{\rm disk}/0.1)(f_{\rm drift}/0.1)$}

In the case of tight binaries, following Jackson et al. (\citeyear{2018MNRAS.477L..85J}), we will assume that: (1) the circumbinary disk mass is a constant fraction of the total system mass, $M_{\rm disk}$ = 0.1$M_{\rm sys}$ (close to the gravitational instability limit); (2) about 10\% of this material migrates in, crossing the unstable radius at which point objects are ejected (where the migration is caused by gas drag, but only operates if planetesimals are relatively small $\lesssim$1 km); and (3) as before, the gas-to-dust ratio is 100:1. This leads to $10^{-4}{\it M_{sys}}(f_{\rm disk}/0.1)(f_{\rm drift}/0.1)$, where we assume $f_{\rm disk}$ = 0.1 and $f_{\rm drift}$ = 0.1. 

\subsection{Fraction of Stars Contributing}
\subsubsection{Singles and Wide Binaries: $f_{\rm pl}$}
\label{FractionContributingSingles}

There is observational evidence that indicates that planetesimal formation is a robust process that can take place under a wide range of conditions: debris disks are present around stars with more than two orders of magnitude difference in stellar luminosity, also in systems with and without binary companions (as circumstellar or circumbinary disks), and around  stars with a wide range of metallicities (unlike planet bearing stars that are strongly correlated with high stellar metallicities Fisher \& Valenti \citeyear{2005ApJ...622.1102F}).  This indicates that planetary systems harboring dust-producing planetesimals are more common than those with giant planets. This would be in agreement with the lack of correlation found between the presence of debris disks and the presence of high-mass planets and of low-mass planets (Moro-Mart\'{\i}n et al. \citeyear{2007ApJ...658.1312M} and \citeyear{2015ApJ...801..143M}, respectively), and with the core accretion models of planet formation, where the planetesimals are the building blocks of planets and the conditions required to form planetesimals are less restricted than those to form larger planets. 

Because of the above considerations, and because for single stars and wide binaries the presence of massive planets seems necessary to provide a mechanism to scatter planetesimals into interstellar space, we will assume that for these stars, the fraction contributing to the population of interstellar planetesimals is determined by those harboring massive planets, $f_{\rm pl}$, rather than those that harbor debris disks (which are evidence of the presence of dust-producing planetesimals but have the caveat that the debris disks surveys are sensitivity-limited). 

For the fraction of stars harboring massive planets we adopt the nominal values of:
\begin{itemize}
\item $f_{\rm pl}$ = 0.2 for A--K2 stars (Marcy et al. \citeyear{2005PThPS.158...24M}; Cumming et al. \citeyear{2008PASP..120..531C}; for A-type stars there are no firm statistics because radial velocity  studies are complicated by the rotational broadening of the absorption lines, the decreased number of spectral features due to high surface temperature, and a large excess velocity resulting from inhomogeneities and pulsation; Johnson et al. \citeyear{2007ApJ...665..785J} estimate that A--F stars are five times more likely than M dwarfs to harbor a giant planet).
\item $f_{\rm pl}$ = 0.03 for K2--M stars (Johnson et al. \citeyear{2007ApJ...670..833J} found that out of 300 M dwarfs monitored, only two have Jupiter-mass planets, while six have Neptune-Uranus mass planets, giving a total frequency of planets around M dwarfs of $\sim$ 3\%; for the purposes of this paper, we will assume $f_{\rm pl}$ = 0.03 because for the lower gravitational potential of low-mass stars, Neptune-mass planets can efficiently eject planetesimals). 
\end{itemize}

\subsubsection{Tight Binaries}
Following Jackson et al. (\citeyear{2018MNRAS.477L..85J}), we assume that all tight binary systems would contribute to the background population of interstellar objects because the ejection mechanism is provided by the tight binary dynamics rather than the presence of a massive planet.  

\subsection{Ejection Efficiency}
\label{ejection}
\subsubsection{Singles and Wide Binaries: ($f_{\rm eject}$/1)}

The dynamical history of the solar system can be relatively well constrained through models and observations and, as discussed in $\S$\ref{Introduction}, these indicate that only a very small fraction of negligible mass of the initial planetesimal disk survived gravitational ejection. As discussed in $\S$1, the efficiency of planetesimal ejection is very sensitive to the dynamical history  of the system, which in turn depends strongly on the planetary configuration and planetesimal disk properties. Even though the large diversity of planetary systems indicates that their dynamical histories are also diverse, it is reasonable to assume that the efficiency of planetesimal ejection is high for many other planetary systems. See for example the dynamical models described in Raymond et al. (\citeyear{2018MNRAS.476.3031R}), that show that there is a high fraction of giant planet systems that are unstable, $\sim$ 90\%, and that these systems are extremely efficient at ejecting planetesimals (most material out to 30 au). We therefore assume that all the circumstellar mass available to form solids is ejected in the form of objects with a wide range of sizes when massive planets are present, and only a negligible fraction is left behind (i.e. $f_{\rm eject}$ = 1).

\subsubsection{Tight Binaries: ($f_{\rm eject}$/1)}

In the case of tight binaries, following Jackson et al. (\citeyear{2018MNRAS.477L..85J}), we assume that all the material that drifts inside the unstable radius (estimated to be 10\% of the total mass) is ejected from the binary system (i.e. $f_{\rm eject}$ = 1). Under this assumption, tight binaries would contribute as much as single stars with planets (per star). 

\subsection{Resulting Mass Density}
\label{ResultingMassDensity}

Using Equations \eqref{mtotalSingles} and \eqref{mtotalBinaries}, we can now calculate the contribution of stars in the different mass ranges considered to the total mass density of interstellar bodies, that would be distributed in objects with a wide range of sizes. The results are shown in Table 1 (column five), showing the dependencies on the different factors assumed. 

\begin{itemize}  
\item For single stars and wide binaries, we estimate that the contributions from the different mass ranges are the following:
\begin{itemize}
\item From M--K2 stars (0.1--0.8 \msol):
\begin{equation}
\label{MassDensitySinglesMK2}
\begin{split}
m_{\rm total}^{\rm singles~M-K2}=\\
1.2\cdot10^{26}\left({{\it f_{\rm solids}} \over 10^{-4}}\right)\left({{\it f_{\rm eject}} \over 1}\right)\left({1-{\it f_{\rm bin}} \over 0.74}\right)\left({\it{f_{\rm pl}} \over \rm 0.03}\right)~{\rm g/pc^{3}},
\end{split}
\end{equation}
\item  From K2--A stars (0.8--2.9 \msol):
\begin{equation}
\label{MassDensitySinglesK2A}
\begin{split}
m_{\rm total}^{\rm singles~K2-A}=\\
5.5\cdot10^{26}\left({{\it f_{\rm solids}} \over 10^{-4}}\right)\left({{\it f_{\rm eject}} \over 1}\right)\left({1-{\it f_{\rm bin}} \over 0.74}\right)\left({\it{f_{\rm pl}} \over \rm 0.2}\right)~{\rm g/pc^{3}}.
\end{split}
\end{equation}
For the latter, we do not consider larger stellar masses because for $M_* >$3 \msol~the snowline moves quickly beyond 10--15 au before protoplanets form, limiting the formation of gas giant planets that under this scenario are required for the ejection of planetesimals. 
\end{itemize}
\item For tight binaries, we estimate that the contribution from systems with masses in the range 0.1--8 \msol~is:
\begin{equation}
\label{MassDensityBinaries}
\begin{split}
m_{\rm total}^{\rm binaries}=\\
2.7\cdot10^{27}({{\it f_{\rm disk}} \over 0.1})({{\it f_{\rm drift} \over 0.1}})({{\it f_{\rm bin}} \over 0.26})~{\rm g/pc^{3}}. 
\end{split}
\end{equation}
In this case, we take into account the contribution of intermediate-mass stars, as the presence of massive planets is not required for the ejection of the material that drifts in and crosses the unstable orbit of the system.
\end{itemize}

The values derived from Equations \eqref{MassDensitySinglesMK2}--\eqref{MassDensityBinaries}  are reflected by the dotted lines in Figures \ref{2s-nom}--\ref{2s-N5E12}, where the top line, labeled  {\it singles (max)}, corresponds to the maximum mass density of interstellar material  expected to originate from single and wide binary systems if we were to assume that all stars, independently on whether or not they host a planet, contribute. The middle line, labeled {\it close binaries}, corresponds to the expected contribution from tight binary systems. And the bottom line, labeled {\it singles (nominal)}, is the contribution expected from single stars and wide binaries in the nominal case under which only stars with giant planets contribute (or Neptune-size planets in the case of low-mass stars). 

\renewcommand\thetable{1}
\LongTables
\begin{deluxetable*}{lllll}
\tablewidth{0pc}
\tablecaption{Expected Mass Density of Ejected Extra-solar Material}
\tablehead{
\colhead{SpType} &
\colhead{Mass Range} &
\colhead{Expected Mass Density} &
\colhead{Fraction of} &
\colhead{Mass Density} \\
\colhead{of Host System} &
\colhead{of Host System} &
\colhead{if all Stars Contributed} &
\colhead{Contributing Stars} &
\colhead{}\\
\colhead{} &
\colhead{(\msol)} &
\colhead{(\msol pc$^{-3}$)} &
\colhead{ } &
\colhead{(g pc$^{-3}$)}
}
\startdata
\multicolumn{3}{l}{Planetesimal disk: single stars and wide binaries:}\\
 & 		&$\int$ $\xi$({\it M$_{*}$})10$^{-4}${\it M$_{*}$}(${{\it f_{\rm solids}} \over 10^{-4}}$)(${{\it f_{\rm eject}} \over 1}$){\it dM$_{*}$} & $(1-{\it f_{\rm bin}})\it{f_{\rm pl}}$ & \\
& & & &\\
M		& 0.1--0.5							& 2.1$\cdot$10$^{-6}$(${{\it f_{\rm solids}} \over 10^{-4}}$)(${{\it f_{\rm eject}} \over 1}$)		& (1-0.26)$\cdot$0.03		& 
9.3$\cdot$10$^{25}$(${{\it f_{\rm solids}} \over 10^{-4}}$)(${{\it f_{\rm eject}} \over 1}$)$({1-{\it f_{\rm bin}} \over 0.74})({\it{f_{\rm pl}} \over \rm 0.03}$)\\
M--K2	& 0.5--0.8							& 6.1$\cdot$10$^{-7}$(${{\it f_{\rm solids}} \over 10^{-4}}$)(${{\it f_{\rm eject}} \over 1}$)		& (1-0.26)$\cdot$0.03		& 
2.7$\cdot$10$^{25}$(${{\it f_{\rm solids}} \over 10^{-4}}$)(${{\it f_{\rm eject}} \over 1}$)$({1-{\it f_{\rm bin}} \over 0.74})({\it{f_{\rm pl}} \over \rm 0.03}$)\\
K2--G	& 0.8--1							& 4.3$\cdot$10$^{-7}$(${{\it f_{\rm solids}} \over 10^{-4}}$)(${{\it f_{\rm eject}} \over 1}$)		& (1-0.26)$\cdot$0.2		& 
1.3$\cdot$10$^{26}$(${{\it f_{\rm solids}} \over 10^{-4}}$)(${{\it f_{\rm eject}} \over 1}$)$({1-{\it f_{\rm bin}} \over 0.74})({\it{f_{\rm pl}} \over \rm 0.2}$)\\
G--F	& 1--1.8						  	& 9.1$\cdot$10$^{-7}$(${{\it f_{\rm solids}} \over 10^{-4}}$)(${{\it f_{\rm eject}} \over 1}$)		& (1-0.26)$\cdot$0.2	 	& 
2.7$\cdot$10$^{26}$(${{\it f_{\rm solids}} \over 10^{-4}}$)(${{\it f_{\rm eject}} \over 1}$)$({1-{\it f_{\rm bin}} \over 0.74})({\it{f_{\rm pl}} \over \rm 0.2}$)\\
A		& 1.8--2.9	& 5.1$\cdot$10$^{-7}$(${{\it f_{\rm solids}} \over 10^{-4}}$)(${{\it f_{\rm eject}} \over 1}$)		&(1-0.26)$\cdot$0.2	 		& 
1.5$\cdot$10$^{26}$(${{\it f_{\rm solids}} \over 10^{-4}}$)(${{\it f_{\rm eject}} \over 1}$)$({1-{\it f_{\rm bin}} \over 0.74})({\it{f_{\rm pl}} \over \rm 0.2}$)\\
M--K2 & 0.1--0.8						& & & 
1.2$\cdot$10$^{26}$(${{\it f_{\rm solids}} \over 10^{-4}}$)(${{\it f_{\rm eject}} \over 1}$)$({1-{\it f_{\rm bin}} \over 0.74})({\it{f_{\rm pl}} \over \rm 0.03}$)\\
K2--A 	& 0.8--2.9							& & & 
5.5$\cdot$10$^{26}$(${{\it f_{\rm solids}} \over 10^{-4}}$)(${{\it f_{\rm eject}} \over 1}$)$({1-{\it f_{\rm bin}} \over 0.74})({\it{f_{\rm pl}} \over \rm 0.2}$)\\
\\
\hline
\\
\multicolumn{3}{l}{Planetesimal disk: close binaries:}\\
 & {\it M$_{\rm sys}$}		&$\int$ $\xi$({\it M$_{\rm sys}$})10$^{-4}${\it M$_{\rm sys}$}(${{\it f_{\rm disk}} \over 0.1}$)(${{\it f_{\rm drift}} \over 0.1}$)(${{\it f_{\rm eject}} \over 1}$){\it dM$_{\rm sys}$} & ${\it f_{\rm bin}}$ & \\
& & & &\\
& 0.1--0.5							& 2.1$\cdot$10$^{-6}$(${{\it f_{\rm disk}} \over 0.1}$)(${{\it f_{\rm drift}} \over 0.1}$)		& 0.26		& 
1.1$\cdot$10$^{27}$(${{\it f_{\rm disk}} \over 0.1}$)(${{\it f_{\rm drift} \over 0.1}}$)$({{\it f_{\rm bin}} \over 0.26})$\\
& 0.5--0.8							& 6.1$\cdot$10$^{-7}$(${{\it f_{\rm disk}} \over 0.1}$)(${{\it f_{\rm drift}} \over 0.1}$)		& 0.26		& 
3.2$\cdot$10$^{26}$(${{\it f_{\rm disk}} \over 0.1}$)(${{\it f_{\rm drift} \over 0.1}}$)$({{\it f_{\rm bin}} \over 0.26})$\\
& 0.8--1							& 4.3$\cdot$10$^{-7}$(${{\it f_{\rm disk}} \over 0.1}$)(${{\it f_{\rm drift}} \over 0.1}$)		& 0.26		& 
2.2$\cdot$10$^{26}$(${{\it f_{\rm disk}} \over 0.1}$)(${{\it f_{\rm drift} \over 0.1}}$)$({{\it f_{\rm bin}} \over 0.26})$\\
& 1--1.8						  	& 9.1$\cdot$10$^{-7}$(${{\it f_{\rm disk}} \over 0.1}$)(${{\it f_{\rm drift}} \over 0.1}$)		& 0.26	 	& 
4.7$\cdot$10$^{26}$(${{\it f_{\rm disk}} \over 0.1}$)(${{\it f_{\rm drift} \over 0.1}}$)$({{\it f_{\rm bin}} \over 0.26})$\\
& 1.8--2.9							& 5.1$\cdot$10$^{-7}$(${{\it f_{\rm disk}} \over 0.1}$)(${{\it f_{\rm drift}} \over 0.1}$)		&0.26		& 
2.6$\cdot$10$^{26}$(${{\it f_{\rm disk}} \over 0.1}$)(${{\it f_{\rm drift} \over 0.1}}$)$({{\it f_{\rm bin}} \over 0.26})$\\
& 2.9--8							& 6.5$\cdot$10$^{-7}$(${{\it f_{\rm disk}} \over 0.1}$)(${{\it f_{\rm drift}} \over 0.1}$)		&0.26 		& 
3.4$\cdot$10$^{26}$(${{\it f_{\rm disk}} \over 0.1}$)(${{\it f_{\rm drift} \over 0.1}}$)$({{\it f_{\rm bin}} \over 0.26})$\\
& 0.1--8	& & & 
2.7$\cdot$10$^{27}$(${{\it f_{\rm disk}} \over 0.1}$)(${{\it f_{\rm drift} \over 0.1}}$)$({{\it f_{\rm bin}} \over 0.26})$\\
\enddata
\end{deluxetable*}

\subsection{Comparison between Inferred and Expected $m_{\rm total}$}
\label{Comparison}

As mentioned above, the hypothesis that we want to test  is whether the source of the population of interstellar objects could be the circumstellar and circumbinary planetesimal disks that form as part of the planetesimal/planet formation process, assuming that 1I/'Oumuamua is representative of a background population of interstellar objects that are isotropically distributed. To test this hypothesis, we now compare the mass density of interstellar objects that would be expected to have a planetesimal disk origin (discussed in $\S$ \ref{ResultingMassDensity}) to the value inferred from the detection of 1I/'Oumuamua (discussed in $\S$ \ref{MassDensity}). 

Given the most likely cometary nature of the object (based on its surface physical properties, Jewitt et al. \citeyear{2017ApJ...850L..36J}; bulk density inferred from its tumbling state, Drahus et al. \citeyear{2018NatAs...2..407D}; and recent evidence of outgassing to account for its nongravitational acceleration, Micheli et al. \citeyear{2018Natur.559..223M}), we will focus the comparison on the contribution from circumstellar planetesimal disks around single stars and wide binaries, as the majority of the objects expected to be ejected from these sources would be icy\footnote{This is because the Safronov number for a planet of a given mass increases with orbital radius, making ejection more efficient beyond the snowline (see e.g. Raymond et al. \citeyear{2018MNRAS.476.3031R}). In addition, debris disk surveys also indicate that Kuiper belts are more common than Asteroid belts (see review in Moro-Mart{\'{\i}}n \citeyear{2013pss3.book..431M}).}.  

An order of magnitude estimate of this contribution can be calculated by adding Equations \eqref{MassDensitySinglesMK2} (M--K2 stars) and \eqref{MassDensitySinglesK2A} (K2--A stars). If we consider the nominal case in which only giant planet-hosts contribute (or Neptune-size planets in the case of low-mass stars), i.e. adopting $f_{\rm pl}$ = 0.03 in Equation \eqref{MassDensitySinglesMK2} and $f_{\rm pl}$ = 0.2 in Equation \eqref{MassDensitySinglesK2A} (because planets are required for the ejection mechanism), this number would be $m_{\rm total}$ =  6.7 $\cdot~10^{26}$ g pc$^{-3}$. We refer to this case as  {\it singles (nominal)}. If we consider the upper limit case in which all stars contribute, i.e. adopting  $f_{\rm pl}$ = 1 in both equations, we get $m_{\rm total}$ =  6.7 $\cdot~10^{27}$ g pc$^{-3}$. We refer to this case as  {\it singles (max)}. This latter case is particularly relevant in the context of M dwarfs, as Kepler results indicate that there is a high incident rate of terrestrial planets around these stars (Dressing \& Charboneau \citeyear{2015ApJ...807...45D}).

In this context, circumbinary disks around tight binary systems could also be a potential source because of icy objects. Jackson et al. (\citeyear{2018MNRAS.477L..85J}) predicts that 64\% of the ejected material in this case would be devolatized, having spent significant time close to the binary stars before being ejected, while the remaining 36\% would be icy. We find that both contributions will add up to a total of  $m_{\rm total}$ =  2.7 $\cdot~10^{27}$ g pc$^{-3}$ [from Equation \eqref {MassDensityBinaries}]. When accounting for the fraction of icy bodies only, this would be reduced to $m_{\rm total}$ =  9.7 $\cdot~10^{26}$ g pc$^{-3}$, slightly larger than the contribution from the {\it singles (nominal)} case described above. 

Figures 1--7 show that, when assuming---for the population of objects from which 1I/'Oumuamua is drawn---a size distribution consisting in a power law with two slopes (discussed in $\S$ \ref{TwoSlopes}) and five slopes (discussed in $\S$ \ref{FiveSlopes}), the comparison between the resulting $m_{\rm total}$ and the estimated values mentioned above (shown as dotted lines in the Figures), leads to the following conclusions: 
\begin{itemize}

\item From Figure \ref{2s-nom}: If we assume the nominal values for the cumulative number density $N_{\rm r \geqslant R}$ = 2 $\cdot~10^{15}$ pc$^{-3}$, and for the object radius $R$ = 80 m, the inferred $m_{\rm total}$ that are closer to the estimated values mentioned above are of the order of $m_{\rm total}$ $\sim$  1 $\cdot~10^{29}$ g pc$^{-3}$. This is about a factor 150 larger than the $m_{\rm total}$ estimated for the {\it singles (nominal)} case, a factor 15 larger than that of the {\it singles (max)} case, and a factor 100 larger than that of the {\it binaries} case, where the best fit models are for ${\it q}_{\rm1}$ $\sim$ 4 and ${\it q}_{\rm2}$ $\gtrsim$ 4. 

\item From Figure \ref{2s-R40}: If we adopt a lower value for the object radius of $R$ = 40 m (if its albedo were to be higher, as some authors have suggested; see discussion in $\S$ \ref{ObjectRadius}), the inferred $m_{\rm total}$ that are closer to the estimated values are still an order of magnitude larger than that of the {\it singles (nominal)} case, a factor of a few larger than that of the {\it singles (max)} case, and an order of magnitude larger than the {\it binaries} case, where the best fit models are for ${\it q}_{\rm1}$ $\sim$ 4 and ${\it q}_{\rm2}$ $\gtrsim$ 3.5--4. Note that some authors have suggested a larger effective radius for 1I/'Oumuamua (e.g. 102 m by Meech et al. \citeyear{2017Natur.552..378M} and 130 m by Bolin et al. \citeyear{2018ApJ...852L...2B}). For these large radii, the inferred $m_{\rm total}$ would diverge even further from the estimated values than when assuming $R$ = 80 m. 

\item From Figure \ref{2s-N5E13}: If we adopt for the cumulative size distribution the value of $N_{\rm r \geqslant R}$ = 5 $\cdot~10^{13}$ pc$^{-3}$ (a lower limit from Feng \& Jones \citeyear{2018ApJ...852L..27F}), keeping $R$ = 80 m, the inferred $m_{\rm total}$ that are closer to the estimated values are almost an order of magnitude larger than that of  the {\it singles (nominal)} case,  close to the value of the {\it singles (max)} case, and a factor of a few larger than the  {\it binaries} case, where the best fit models are for ${\it q}_{\rm1}$ $\gtrsim$ 3.5 and ${\it q}_{\rm2}$ $\gtrsim$ 3.5. But this lower value of $N_{\rm r \geqslant R}$, a factor of 30 lower than the initial assumption of 2 $\cdot~10^{15}$ pc$^{-3}$ from Do et al. (\citeyear{2018ApJ...855L..10D}), does not seem to be favored by their detailed study of the survey volume, and might be outside the uncertainty range for the cumulative number density\footnote{Do et al. (\citeyear{2018ApJ...855L..10D}) note that the cumulative number density of 2 $\cdot~10^{15}$ pc$^{-3}$ is an underestimate of at most 40\% because the objects have a cumulative size distribution that falls at larger sizes; they also point out that the detection process is not 100\% efficient and, given these inefficiencies, the detection volume could be 2/3--3/4 of the nominal value and therefore the number density could be 4/3--3/2 of their inferred number density.}.

\item From Figure \ref{2s-R40N5E13}: When adopting $N_{\rm r \geqslant R}$ = 5 $\cdot~10^{13}$ pc$^{-3}$ and $R$ = 40 m, the inferred $m_{\rm total}$ that are closer to the estimated values would be close to that of the {\it singles (nominal)} case and the  {\it binaries} case, where the best fit models are for ${\it q}_{\rm1}$ $\sim$ 4 and ${\it q}_{\rm2}$ $\gtrsim$ 3.5. 

\item From Figure \ref{2s-N5E12}:  Exploring what value of the cumulative number density would lead, for some of the size distributions considered,  and assuming $R$ = 80 m, to a mass density $m_{\rm total}$ that is in agreement with that expected from planetesimals disks in the {\it singles (nominal)} case, we get $N_{\rm r \geqslant R}$ $\sim$ 5$\cdot$10$^{12}$ pc$^{-3}$. For the {\it singles (max)} case, this value would be $N_{\rm r \geqslant R}$ $\sim$ 5$\cdot$10$^{13}$ pc$^{-3}$. From these values, and scaling from the observed frequency of detections for PanSTARRS (1 in 3.5 years for $N_{\rm r \geqslant R}$ $\sim$ 2$\cdot$10$^{15}$ pc$^{-3}$), we can estimate what the PanSTARRS detections per year would be that we would expect from the estimated population of objects with a circumstellar disk origin. These frequencies would be: 1 detection every $\sim$1000 years for the {\it singles (nominal)} case, a slightly higher frequency for the  {\it binaries} case, and 1 detection every $\sim$100 years for the {\it singles (max)} case. 

\begin{figure}
\begin{center}
\includegraphics[width=6cm]{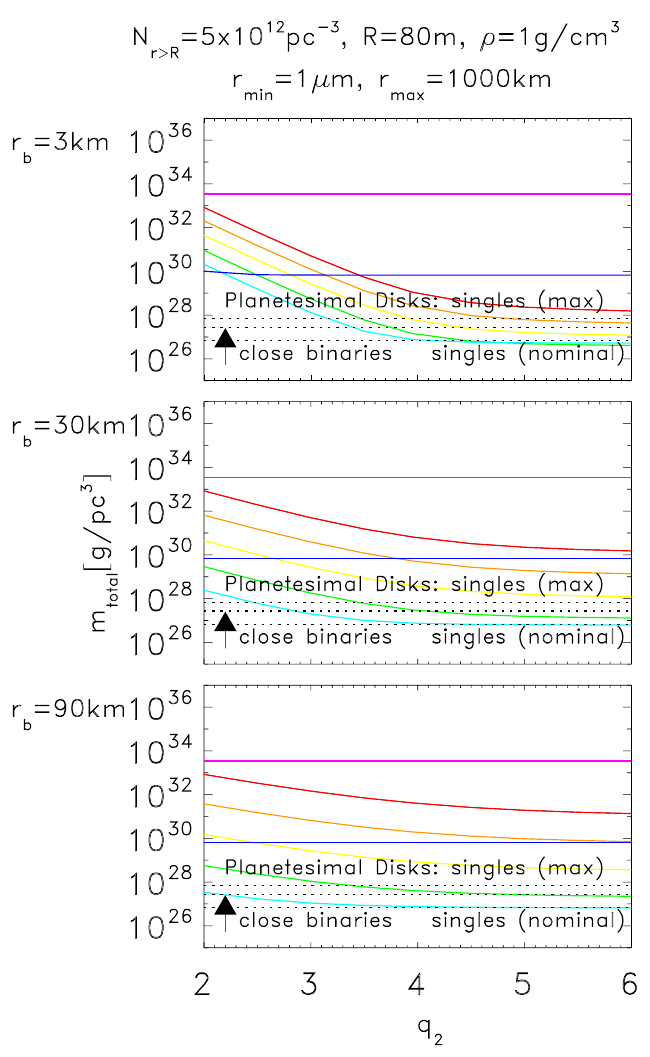}
\end{center}
\caption{Same as Figure \ref{2s-nom} but exploring what value of the cumulative number density, $N_{\rm r \geqslant R}$, would lead for some of the size distributions considered, to a mass density in agreement with that expected from planetesimals disks in the {\it singles (nominal)} case (in which only stars with planets contribute), assuming the object size is 80 m. This value is $\sim$ 5 $\cdot10^{12}$ pc$^{-3}$. 
}
\label{2s-N5E12}
\end{figure}

\item From Figure \ref{5s-4G}: If we assume the size distribution in the five-slope case corresponding to the solar system at 4.5 Gyr, and adopting the nominal values of $N_{\rm r \geqslant R}$ = 2 $\cdot~10^{15}$ pc$^{-3}$ and $R$ = 80 m, we get $m_{\rm total}$ =  2.4 $\cdot~10^{30}$ g pc$^{-3}$, about a factor of 3600 larger than the {\it singles (nominal)} case, 360 larger than the {\it singles (max)} case, and 890 larger than the  {\it binaries} case. If we adopt a lower value of $R$ = 40 m, we get $m_{\rm total}$ =  5.0 $\cdot~10^{29}$ g pc$^{-3}$, about a factor of 750 larger than the {\it singles (nominal)} case,  75 larger than the {\it singles (max)} case, and 190 larger than the  {\it binaries} case. Assuming $R$ = 80 m, exploring what value of the cumulative number density would lead to a mass density $m_{\rm total}$ that is in agreement with that expected from planetesimals disks in the {\it singles (nominal)} case, we get $N_{\rm r \geqslant R}$ $\sim$ 3.3$\cdot$10$^{11}$ pc$^{-3}$, while for the {\it singles (max)} case, this value would be $N_{\rm r \geqslant R}$ $\sim$ 4$\cdot$10$^{12}$ pc$^{-3}$. Scaling from the observed frequency of detections for PanSTARRS, the corresponding detection frequencies would be: 1 detection every $\sim$18000 years for the {\it singles (nominal)} case, and 1 detection every $\sim$1400 years for the {\it singles (max)} case. 

\item From Figure \ref{5s-100M}: If we assume the size distribution in the five-slope case corresponding to the solar system at 100 Myr, and adopting the nominal values of $N_{\rm r \geqslant R}$ = 2 $\cdot~10^{15}$ pc$^{-3}$ and $R$ = 80 m, we get $m_{\rm total}$ =  7.4 $\cdot~10^{29}$ g pc$^{-3}$, about a factor of 1100 larger than the {\it singles (nominal)} case, 110 larger than the {\it singles (max)} case, and 270 larger than the {\it binaries} case. If we adopt a lower value of $R$ = 40 m, we get $m_{\rm total}$ =  1.5 $\cdot~10^{29}$ g pc$^{-3}$,   about a factor of 220 larger than the {\it singles (nominal)} case, 22 larger than the {\it singles (max)} case, and 55 larger than the {\it binaries} case. The estimates regarding the value of the cumulative number density that would lead,  assuming $R$ = 80 m, to a mass density $m_{\rm total}$ that is in agreement with that expected from planetesimals disks in the {\it singles (nominal)} case is $N_{\rm r \geqslant R}$ $\sim$ 6.3$\cdot$10$^{11}$ pc$^{-3}$, while for the {\it singles (max)} case this value would be $N_{\rm r \geqslant R}$ $\sim$ 1.6$\cdot$10$^{13}$ pc$^{-3}$. This would correspond to 1 PanSTARRS detection every $\sim$9000 years for the {\it singles (nominal)} case, and 1 every $\sim$400 years for the {\it singles (max)} case. 

\end{itemize}

\section{Implications Regarding 1I/'Oumuamua's Origin}
\label{ImplicationsOrigin}

\subsection{Planetesimal Disk Origin}
\label{PlanetesimalDisk}
The conclusion based on our analysis above is that 1I/'Oumuamua is unlikely to be representative of an isotropically distributed background population that would result from the ejection of planetesimals that form in circumstellar and circumbinary disks, and that are part of a collisional population. This is because, if we assume 1I/'Oumuamua represents a distribution that is isotropically distributed and has a wide range of size distributions, the inferred mass density of the reservoir that 1I/'Oumuamua would have been drawn from (adopting the cumulative number density, object radius and bulk density inferred from its observations) would be higher than what could be accounted for by circumstellar and circumbinary planetesimal disks (as discussed in $\S$ \ref{Comparison}).  This is in agreement with Raymond et al. (\citeyear{2018MNRAS.476.3031R}), who reached a similar conclusion with regard to circumstellar disks around single stars, with the difference that our study explores a wide range of size distributions, rather than a single power law, and also addresses the contribution from binary systems. 

The comparison is even less favorable when we take into account that our estimate of the total mass density of interstellar objects ejected from planetesimal disks in single and wide binary systems (described in $\S $\ref{ResultingMassDensity}) is likely to be an overestimate. This is because it assumes that most of the material around the stars harboring giant planets is ejected (as discussed in $\S$ \ref{FractionContributingSingles}).  Ejection is indeed efficient around 1 M$_{\rm Jupiter}$ planets located at 1--30 au, 0.1--10 M$_{\rm Jupiter}$ planets at $\sim$5 au, Saturn-mass planets at 10-30 au, eccentric planets and long-period giant planets (Wyatt et al. \citeyear{2017MNRAS.464.3385W}), but this is for the material that crosses the orbits of these planets. 
 
A scenario that has been proposed is that 1I/'Oumuamua originated from the planetesimal disk of a young nearby star (Gaidos et al. \citeyear{2017RNAAS...1a..13G}; Gaidos \citeyear{2018MNRAS.477.5692G}), in which case the ejected bodies would not be isotropically distributed. This highly anisotropic distribution, resulting in large fluctuations in space density, may favor their detection if the solar system is in its path, leading to an overestimate of the background number density, assuming it is isotropically distributed (a critical assumption we make in our study). 
 
There are several observations that might support a young age. One is the color of the object, found not to be as red as the ultra-red bodies in the outer solar system (Jewitt et al. \citeyear{2017ApJ...850L..36J}), thought to be reddened by space weathering (from cosmic rays and ISM plasma). This suggests that 1I/'Oumuamua has not traveled in interstellar space during Gyr (Gaidos et al. \citeyear{2017RNAAS...1a..13G}; Feng \& Jones \citeyear{2018ApJ...852L..27F}; Fitzsimmons et al. \citeyear{2018NatAs...2..133F}). 

A weaker indication of youth is its tumbling state, thought to have originated in a collision (because, if arising from sublimation torques, the sublimation levels required would have resulted in a rotational rate high enough to lead to to the rotational destruction of the object). And most probably, this collision (which could be a catastrophic collision, or collisions with minor bodies) happened at its home system (because collisions in the solar system, in particular given its highly inclined orbit, would be unlikely and would have most likely left a dust trail; Drahus et al. \citeyear{2018NatAs...2..407D}). Drahus et al. (\citeyear{2018NatAs...2..407D}) estimates that, for a rubble pile object with a size, shape, density, and rotational rate as inferred from 1I/'Oumuamua's observations,  the damping time scale of the rotational energy (due to stresses and strains that result from its complex rotation)  would be $\sim$ 1 Gyr, implying that the object is younger than that. But these damping times are uncertain, as Fraser et al. (\citeyear{2018NatAs...2..383F}) estimate a damping time of 3$\cdot$10$^{9}$ -- 3$\cdot$10$^{11}$ years for icy objects like 1I/'Oumuamua.  The tumbling state would therefore be consistent with the scenario described above, where the object originated in a planetesimal disk around a young star because, during these periods of dynamical instability, frequent collisions are common  (like during the Late Heavy Bombardment in the solar system).
  
 The strongest argument supporting youth comes from the kinematics of the object, as it is estimated that its velocity before it entered the solar system was within 3--10 \kms of the velocity of the {\it LSR} (Gaidos et al. \citeyear{2017RNAAS...1a..13G}; Mamajek \citeyear{2017RNAAS...1a..21M}, Do et al. \citeyear{2018ApJ...855L..10D}), like many young stellar associations. This is because, with time, the dynamics of the objects would have been altered by passing stars, clouds, spiral arms, and star clusters, perturbing this initial LSR-like motion, resulting in a larger relative velocity with respect to the LSR. Based on this argument, Gaidos et al. (\citeyear{2017RNAAS...1a..13G}) estimate an age of $\ll$1 Gyr, while Feng \& Jones (\citeyear{2018ApJ...852L..27F}) indicate that the probability of observing the object with a velocity $<$10 \kms with respect to the LSR is 0.5, 0.26, and 0.13, for ages of 0.1 Gyr, 1 Gyr, and 10 Gyr, respectively (and these probabilities might be smaller for low-mass objects like 1I/'Oumuamua, as their orbits are more easily altered, further supporting a young age). Gaidos (\citeyear{2018MNRAS.477.5692G}) estimates that, if originating from the overall stellar population, the probability that the velocity of 1I/'Oumuamua coincides with that of a young association is $<$1\%. The kinematics of the object therefore indicates a recent origin of ejection from the planetesimal disk of a young nearby association ($<$ 100 Myr). 

Some authors have attempted to identify the star or association from which 1I/'Oumuamua originated, but these studies generally do not take into account the errors in stellar positions, which is the most important source of uncertainty (Dybczy{\'n}ski \& Kr{\'o}likowska \citeyear{2018A&A...610L..11D}). Other caveats are that young stellar associations are dispersed over 10's of pc (Gaidos \citeyear{2018MNRAS.477.5692G}), and that given the typical interstellar distance between encounters, large perpendicular displacements can be produced, making it difficult to predict the result of successive stellar encounters (Zhang \citeyear{2018ApJ...852L..13Z}).  

\subsection{Other Proposed Origins}
\label{OtherOrigins}

Another origin that has been proposed is ejection from a white dwarf system, either by direct ejection (Rafikov \citeyear{2018ApJ...861...35R}), or in a  tidal disruption event (Hansen \& Zuckerman \citeyear{2017RNAAS...1a..55H}). These objects, scattered by massive planets into low periastron orbits, have been proposed to account for the observed white dwarf atmospheric pollution.  But the main motivation underlying the study of a potential white dwarf origin was the assumed refractory nature of 1I/'Oumuamua (based on its lack of cometary activity), as the objects around the white dwarfs would have been devolatized during the giant phase of the central star. The recently found nongravitational acceleration of the object, and the fact that it is consistent with outgassing (Micheli et al. \citeyear{2018Natur.559..223M}), together with other considerations,  favor an icy composition, making a white dwarf origin less likely.  Another caveat of this scenario is that the expected population of remnants would exhibit kinematic characteristics similar to that of old stars, as they would have experienced dynamical heating in the galaxy by gravitational scattering with massive objects, contrary to the observed kinematic properties of 1I/'Oumuamua (as described in $\S$ \ref{PlanetesimalDisk}). 

Other sources that have been proposed, related to compact sources, are: (1) A fragment of a tidally disrupted object around a compact star in a binary system ({\'C}uk \citeyear{2018ApJ...852L..15C}); but this assumes that the material is channeled into 100 m-sized objects, and also requires a parameter space in terms of binary properties that might be limited. (2) It was an object originally orbiting a star than underwent a core collapse supernova explosion and that was unbound after the sudden loss of stellar mass; but this scenario would not be able to account for the observation that 1I/'Oumuamua is dynamically cold and that the objects might have become devolatized or not survived the blast (Rafikov \citeyear{2018ApJ...861...35R}). 

Of our two basic assumptions, that the object is representative of an isotropically distributed population, and that it is the result of a collisional cascade, we have addressed the possibility that the object does not fulfill the former (see $\S$ \ref{PlanetesimalDisk}). 

But it could also be the case that the object might not be the result of a collisional cascade. Rafikov (\citeyear{2018ApJ...861...35R}) argues that, in the scenario of tidal disruption around a white dwarf, collisional fragmentation could channel most of the original material into 0.1--1 km sized bodies, encompassing 1I/'Oumuamua's size, because this is the size at which the collisional velocity leading to catastrophic disruption is minimized. As pointed out above, {\'C}uk (\citeyear{2018ApJ...852L..15C}) also argued that the tidally disrupted material would be channeled into 100 m-sized objects. These are interesting hypotheses but we have already pointed out that the icy nature of 1I/'Oumuamua represents a challenge to both scenarios. Raymond et al. (\citeyear{2018ApJ...856L...7R}, \citeyear{2018MNRAS.476.3031R}) also considered the scenario in which 1I/'Oumuamua is a fragment of a tidally disrupted planetesimal, where a  small percent of planetesimals fragment into objects 100 m in size, which would dominate the distribution in number. In fact, their dynamical simulations show that $\sim$1\% of planetesimals pass within the tidal disruption radius of a gas giant on their pathway to ejection. But what is still missing from all these scenarios are the models that predict such a narrow size distribution of the fragments. 

In the context of this discussion, it is of interest  to point out that, according to the models in Schlichting et al. (\citeyear{2013AJ....146...36S}), the dominant sizes from which planets grow might range from a few hundred meters to a few km, encompassing 1I/'Oumuamua's size. So this object could be representative of that primordial population of building blocks. On the other hand, Johansen \& Lambrechts (\citeyear{2017AREPS..45..359J}) indicate that the characteristic scales of planetesimals resulting from streaming instability might be $\sim$100 km, much larger than 1I/'Oumuamua, but these models are not yet able to predict the size distribution resulting from this process at the small end (including 1I/'Oumuamua's size). The size distribution of this initial population of building blocks remains as one of the major open questions in planet formation and could be relevant in the context of the interstellar interlopers. In the case of the solar system, this initial size distribution has left a permanent signature in the current size distribution of small KBOs, so the study of the latter at smaller and smaller sizes will help to constrain it (Schlichting et al. \citeyear{2013AJ....146...36S}). 

Future detections of incoming interstellar objects like 1I/'Oumuamua will shed light on the size distribution of this population, which cannot be inferred from the detection of a single object, and this will help address the open questions about their origin.  

 

 

\section{Expected Flux of Interstellar Meteorites and Micrometeorites on Earth}
\label{meteorites}

Now we address whether the flux of meteorites and micrometeorites expected on Earth could be a discriminating measurement regarding the origin of interstellar interlopers. This flux is also of interest because of the possibility, suggested by Gaidos (\citeyear{2018MNRAS.477.5692G}), that one of these interstellar objects may already be part of the collected meteorite samples. In addition, given the unusually high impact velocity of these interlopers compared to other incoming bodies, this flux can assess to what degree their larger counterparts might pose a threat. 

In the calculations below, we consider two size ranges: micrometeorites with diameters in the range 20 $\mu$m--1 mm and meteorites with diameters of 2.7--12.4 cm (corresponding to 10 g--1 kg, if assuming spherical grains with bulk density of 1 g cm$^{-3}$). 

\subsection{Inferred Interstellar Flux from 1I/'Oumuamua's Detection and Estimated Flux from a Planetesimal Disk Origin}

We compare the expected fluxes of meteorites and micrometeorites on Earth inferred from 1I/'Oumuamua's detection to those expected from a planetesimal disk origin, assuming in both cases that the remnants are isotropically distributed and originate from a collisional cascade.

For a given number density of objects, $n$, the average time between impacts at the at the top of the atmosphere is given by, $\Delta t$ = ${\rm 1 \over n\cdot A \cdot v_{0}}$, where $A$ is the Earth's effective cross section, 

\begin{equation} 
\begin{split}
A = \pi R_{\oplus}^2 \left(1+\left(v_{\rm esc} \over v_{0}\right)^2\right)\\
= 1.5\cdot10^{18}~cm^2, 
\end{split}
\end{equation}
and $v_{0}$ = 26 \kms is the velocity of 1I/'Oumuamua with respect to the Earth, $v_{\rm esc}$ = 11.2 \kms is the escape velocity from the Earth, and $R_{\oplus}$ is the Earth's radius. 

To estimate the number density of objects in the two size ranges considered (for meteorites and micrometeorites, respectively), we use the five-slope size distributions derived by Schlichting et al. (\citeyear{2013AJ....146...36S}), given in Equation \eqref{5s-sizedist}, characteristic of the solar system at 100 Myr and at 4.5 Gyr. The number density would be given by 

\begin{equation}
\begin{split}
{n = \int\limits_{R_{min}}^{R_{max}}A'_1r^{-q_1}dr =}\\
{{A'_1\over-q_{1}+1}(R_{max}^{-q_1+1}-R_{min}^{-q_1+1})},
\end{split}
\end{equation}
where $R_{min}$ and $R_{max}$ are the size boundaries considered, and $A'_1$ is given by Equation \eqref{5s-Aip} and depends on the cumulative number density $N_{\rm r \geqslant R}$. 

We now carry out this calculation using the values of $N_{\rm r \geqslant R}$ that correspond to:
\begin{itemize}

\item The cumulative number density inferred from 1I/'Oumuamua's detection (assuming an isotropic distribution), $N_{\rm r \geqslant R}$ = 2 $\cdot10^{15}$ pc$^{-3}$. 

\item The cumulative number density that would originate from the objects ejected from circumstellar planetesimal disks (nominal case), $N_{\rm r \geqslant R}$ = 3.3$\cdot10^{11}$ pc$^{-3}$, when assuming the size distribution at 4.5 Gyr, and $N_{\rm r \geqslant R}$ = 6.3$\cdot10^{11}$ pc$^{-3}$, when assuming the size distribution at 100 Myr (inferred from Figures \ref{5s-4G} and \ref{5s-100M} in $\S$ \ref{Comparison}, respectively). 

\item The cumulative number density that would originate from the objects ejected from circumbinary planetesimal disks, $N_{\rm r \geqslant R}$ = 1.4$\cdot10^{12}$ pc$^{-3}$, when assuming the size distribution at 4.5 Gyr, and $N_{\rm r \geqslant R}$ = 5.6$\cdot10^{12}$ pc$^{-3}$, when assuming the size distribution at 100 Myr (also inferred from Figures \ref{5s-4G} and \ref{5s-100M}). 

\end{itemize}
The resulting values for the expected impact frequencies and  mass fluxes at the top of Earth's atmosphere are listed in Table 2, for both micrometeorites and meteorites.

Comparing the fluxes above, for the 4.5 Gyr case, we find that the mass fluxes derived from 1I/'Oumuamua's detection for both micrometeorites and meteorites exceeds those expected from planetesimal disks by a factor of $\sim$ 6000 in the {\it singles (nominal)} case and a factor of $\sim$ 1400 in the {\it binaries} case. For the 100 Myr case, these factors are $\sim$ 3200 and $\sim$ 360, respectively. 

\renewcommand\thetable{2}
\LongTables
\begin{deluxetable*}{llllllll}
\tablewidth{0pc}
\tablecaption{Micrometeorite and Meteorite Fluxes expected and measured at the top of Earth's atmosphere}
\tablehead{
\colhead{~~~~~~~~~~~~} &
\colhead{'Oumuamua's} &
\colhead{'Oumuamua's} &
\colhead{Planet. disks} &
\colhead{Planet. disks} &
\colhead{Planet. disks} &
\colhead{Planet. disks} &
\colhead{Observed}\\
\colhead{} &
\colhead{derived $N_{\rm r \geqslant R}$} &
\colhead{derived $N_{\rm r \geqslant R}$} &
\colhead{(singles nom.)} &
\colhead{(binaries)} &
\colhead{(singles nom.)} &
\colhead{(binaries)} &
\colhead{at Earth}\\
\colhead{} &
\colhead{(4.5 Gyr)} &
\colhead{(100 Myr)} &
\colhead{(4.5 Gyr)} &
\colhead{(4.5 Gyr)} &
\colhead{(100 Myr)} &
\colhead{(100 Myr)} &
\colhead{}
}
\startdata
$N_{\rm r \geqslant R}$	& 2 $\cdot~10^{15}$	& 2 $\cdot~10^{15}$	& 3.3$\cdot10^{11}$		&	1.4$\cdot10^{12}$	&	6.3$\cdot10^{11}$	&	5.6$\cdot10^{12}$	&												\\
(pc$^{-3}$)\\
\multicolumn{8}{c}{Micrometeorites:}\\
\\
Impact freq.				& 2.6$\cdot10^{10}$	& 2.7$\cdot10^{10}$	&4.2$\cdot10^{6}$		&	1.8$\cdot10^{7}$		&	8.5$\cdot10^{6}$		&	7.5$\cdot10^{7}$		&												\\
(yr$^{-1}$)\\
Mass flux						& 6.0		&	6.2	&	9.7$\cdot10^{-4}$ &	4.1$\cdot10^{-3}$&	1.9$\cdot10^{-3}$	&	1.7$\cdot10^{-2}$	&	(11$\pm$5.5)$\cdot10^{7}$				\\
(g day$^{-1}$)\\
\multicolumn{8}{c}{Meteorites:}\\
\\
Impact freq.				& 89	& 93 & 1.5$\cdot10^{-2}$	&	6.3$\cdot10^{-2}$	&	2.9$\cdot10^{-2}$	&	0.26	&												\\
(yr$^{-1}$)\\
Mass flux 						& 13	& 14	& 2.2$\cdot10^{-3}$		&	9.4$\cdot10^{-3}$	&	4.4$\cdot10^{-3}$	&	3.9$\cdot10^{-2}$	&	(2.9--7.3)$\cdot10^{6}$					\\
(g day$^{-1}$)\\
\enddata
\end{deluxetable*}

\subsection{Observed Flux on Earth}

We now compare the above mass fluxes to those measured at the Earth's atmosphere. Micrometeorites in the present day represent the most important source of extraterrestrial matter that falls to Earth, also contributing importantly to the composition in the regolith of other solar system bodies. Plane et al. (\citeyear{2018SSRv..214...23P}) and references therein estimate that the current influx of dust at the top of the atmosphere is $\sim$ (11$\pm$5.5)$\cdot$10$^{7}$ g day$^{-1}$, of which 80\% would correspond to micrometeorites in approximately the size range considered. For meteorites, Bland et al. (\citeyear{1996MNRAS.283..551B}) estimates a present impact flux on Earth of 2.9--7.3$\cdot$10$^{6}$ g day$^{-1}$ for the mass range of 10 g--1 kg (corresponding to the size range considered in our calculation, when assuming spherical grains and a bulk density of 1 g cm$^{-3}$). 

Because these observed fluxes for meteorites and micrometeorites are many orders of magnitude larger than those inferred from 1I/'Oumuamua's detection and expected from a planetesimal disk origin (see Table 2), these measurements on Earth cannot be used as a discriminating factor. The comparison to the observed fluxes also indicates that it is unlikely that one of these objects with an interstellar origin is already part of the collected meteorite samples, contrary to Gaidos (\citeyear{2018MNRAS.477.5692G}). We also find that these interstellar impactors, even though they have an unusually high impact velocity compared to other incoming bodies, may not pose a particularly high threat because of their very low impact frequency compared to meteorites from other sources.

\section{Conclusions}
\label{Conclusions}

In this paper, we test the hypothesis that 1I/'Oumuamua is representative of a background population of interstellar objects that are isotropically distributed, are the result of a collisional cascade, and were ejected from circumstellar and circumbinary planetesimal disks as a natural product of the planetesimal/planet formation process. The recent observations supporting an icy composition of 1I/'Oumuamua make this comparison very relevant because the majority of these objects (in the case of the single stars and the wide binaries) would have originated beyond the snowline. The interesting factor is that it could provide information about the building blocks of planets in a size range that remains elusive to observations, and this can help to constrain planet and planetesimal formation models.  

We do the test by comparing the mass density of interstellar objects inferred from the detection of 1I/'Oumuamua to that expected from planetesimal disks under two scenarios: circumstellar disks around single stars and wide binaries, and circumbinary disks around tight binaries. Our approach makes use of a detailed study of the PanSTARRS survey volume; takes into account that the contribution from each star to the population of interstellar planetesimals depends on the mass of the central mass, whether the star is in a single, wide binary or tight binary system, and whether it is a planet-host; and takes into account a wide range of possible size distributions for the ejected planetesimals, based on solar system models and observations of its small-body population. 

We find that, if we assume 1I/'Oumuamua represents a collisional population that is isotropically distributed, the inferred mass density of the reservoir that 1I/'Oumuamua would have been drawn from (adopting the cumulative number density, object radius, and bulk density inferred from its observations, and assuming a wide range of size distributions), would be higher than what could be accounted for by circumstellar and circumbinary planetesimal disks. We therefore conclude that 1I/'Oumuamua is unlikely to be representative of such a population.  

We favor the scenario in which 1I/'Oumuamua is a member of a population that is highly anisotropic, which has led to an overestimate of the background number density of objects. This would be in agreement with a scenario that has been proposed where 1I/'Oumuamua originated from the planetesimal disk of a young ($<$100 Myr) nearby star, explaining several of the observations supporting its kinematics, icy composition, and tumbling state. 

Another possibility is that 1I/'Oumuamua is not the result of a collisional cascade but of the tidal disruption of a planetesimal, in a process that would have channeled the majority of the objects into fragments of 1I/'Oumuamua's size. But the resulting size distribution from such a process is still uncertain. Or it could be representative of the primordial population of building blocks from which planets grow, with dominant sizes thought to range from a few hundred meters to a few km, encompassing 1I/'Oumuamua's size. 

Finally, we compare the observed flux of meteorites and micrometeorites on Earth to those inferred from 1I/'Oumuamua's detection, and to those expected from a planetesimal disk origin. We find that in both cases the observed fluxes of meteorites and micrometeorites are many orders of magnitude larger and therefore we conclude it is unlikely that one of these objects are already part of the collected meteorite samples.  

Future detections of incoming interstellar objects like 1I/'Oumuamua will shed light on the size distribution of this population, which cannot be inferred from the detection of a single object, and this will help address the open questions about their origin and meteoritic contribution.  

A.M.-M. thanks David Jewitt for useful discussions and the anonymous referee for helpful suggestions. 

\begin{center}
{\bf APPENDIX}
\end{center}
\begin{center}
{\bf DERIVATION OF THE MASS DENSITY}
\end{center}

{\bf  A.1 Two-slope Size Distribution:}\\

Adopting for the size distribution of interstellar objects a broken power law with two slopes, such as that in Equation \eqref{2s-sizedist}, we calculate the total mass density of interstellar objects for a given number density of objects with radius $r \geqslant R$, and assuming an isotropic spatial distribution.  Under these assumptions, the mass distribution would be,
\begin{equation}
\begin{split}
n(m) = A_1 \cdot m^{-\alpha_1}~~{\rm for}~m < m_b\\
n(m) = A_2 \cdot m^{-\alpha_2}~~{\rm for}~m > m_b,
\end{split}
\label{eq_size_dist_m}
\end{equation} 
where $A_1$ and $A_2$ are calculated from 
continuity, $A_1m_b^{-\alpha_1} = A_2m_b^{-\alpha_2}$, 
and from normalizing to the expected mass density over the full mass range, 
$m_{total} = \int\limits_{m_{min}}^{m_{b}}A_1m^{-\alpha_1}mdm+ \int\limits_{m_{b}}^{m_{max}}A_2m^{-\alpha_2}mdm$, yielding
\begin{equation}
\begin{split}
{A_1 = {m_{total} \over 
m_b^{-\alpha_1+\alpha_2} {m_{max}^{-\alpha_2+2}-m_{b}^{-\alpha_2+2} \over -\alpha_2+2}+
{m_{b}^{-\alpha_1+2}-m_{min}^{-\alpha_1+2} \over -\alpha_1+2}}}\\
{A_2 = {m_{total} \over 
{m_{max}^{-\alpha_2+2}-m_{b}^{-\alpha_2+2} \over -\alpha_2+2}+
m_b^{-\alpha_2+\alpha_1}{m_{b}^{-\alpha_1+2}-m_{min}^{-\alpha_1+2} \over -\alpha_1+2}}}.
\end{split}
\label{A1_A2}
\end{equation}
Assuming all interstellar objects have the same bulk density, $\rho$, the above mass distribution is equivalent to a size distribution of  
\begin{equation}
\begin{split}
n(r) = A'_1 \cdot r^{-q_1}~~{\rm for}~r < r_b\\ 
n(r) = A'_2 \cdot r^{-q_2}~~{\rm for}~r > r_b,
\end{split}
\label{eq_size_dist_r}
\end{equation} 
where $A'_1$, $A'_2$, $q_1$ and $q_2$ are derived from 
$n(m)dm=n(r)dr$ and given by
\begin{equation}
\begin{split}
{A'_1 = 3^{\alpha_1}(4\pi\rho)^{-\alpha_1+1}A_1},\\
{A'_2= 3^{\alpha_2}(4\pi\rho)^{-\alpha_2+1}A_2}, 
\end{split}
\label{A1p_A2p}
\end{equation}
with $q_1 = 3\alpha_1-2$ and $q_2 = 3\alpha_2-2$.
Substituting the expressions in Equation \eqref{A1_A2} into \eqref{A1p_A2p}, and using $m_i = {4 \over 3}\pi\rho r_i^3$ and $\alpha_i = { q_i+2 \over 3}$, we get
\begin{equation}
\begin{split}
{A'_1 = {9(4\pi\rho)^{-1}m_{total}r_b^{q_1-4} \over {3 \over 4-q_2}[({r_{max} \over r_{b}})^{4-q_2}-1]+{3 \over 4-q_1}[1-({r_{min} \over r_{b}})^{4-q_1}]}}\\
{A'_2 = {9(4\pi\rho)^{-1}m_{total}r_b^{q_2-4} \over {3 \over 4-q_2}[({r_{max} \over r_{b}})^{4-q_2}-1]+{3 \over 4-q_1}[1-({r_{min} \over r_{b}})^{4-q_1}]}}.
\end{split}
\label{A1p_A2p_detailed}
\end{equation}
The total number density of interstellar objects with radius $r \geqslant R$~($< r_b$)  is given by $N_{\rm r \geqslant R} = \int\limits_{R}^{r_{b}}A'_1r^{-q_1}dr+ \int\limits_{r_{b}}^{r_{max}}A'_2r^{-q_2}dr$, and from the expressions in \eqref{A1p_A2p_detailed} we get
\begin{equation}
\begin{split}
{N_{\rm r \geqslant R} = {9(4\pi\rho)^{-1}m_{total} \over {3 \over 4-q_2}[({r_{max} \over r_{b}})^{4-q_2}-1]+{3 \over 4-q_1}[1-({r_{min} \over r_{b}})^{4-q_1}]}}\\
\times \left( {1 \over 1-q_1}[{1 \over r_b^{3}}-{1 \over R^{3}}({R \over r_b})^{4-q_1}]+{1 \over 1-q_2}[{1 \over r_{max}^{3}}({r_{max} \over r_b})^{4-q_2}-{1 \over r_b^{3}}]\right)
\end{split}
\label{N_r_gt_R}
\end{equation}
For a given $N_{\rm r \geqslant R}$ in units of pc$^{-3}$, and with $\rho$ in (g cm$^{-3}$) and $R$ and $r_{i}$ in cm, the total mass density of solids, $m_{\rm total}$ (g pc$^{-3}$), would be given by
\begin{equation}
\begin{split}
{m_{total} = {N_{\rm r \geqslant R} \over 9(4\pi\rho)^{-1}}\times}\\
{{3 \over 4-q_2}\left[\left({r_{max} \over r_{b}}\right)^{4-q_2}-1\right]+{3 \over 4-q_1}\left[1-\left({r_{min} \over r_{b}}\right)^{4-q_1}\right]\over \times \left( {1 \over 1-q_1}[{1 \over r_b^{3}}-{1 \over R^{3}}({R \over r_b})^{4-q_1}]+{1 \over 1-q_2}[{1 \over r_{max}^{3}}({r_{max} \over r_b})^{4-q_2}-{1 \over r_b^{3}}]\right)}.
\end{split}
\label{2s-mtotal}
\end{equation}


{\bf  A.2 Five-slope Size Distribution:}\\

Using the the same procedure as above, we now estimate $m_{\rm total}$ for a size distribution approximated as a broken power law with five slopes, such as that in Equation \eqref{5s-sizedist}. In this case, the mass distribution would be given by
\begin{equation}
\begin{split}
n(m) = A_1 \cdot m^{-\alpha_1}~~{\rm for}~m < m_{b1}\\
n(m) = A_2 \cdot m^{-\alpha_2}~~{\rm for}~m_{b1} < m < m_{b2}\\
n(m) = A_3 \cdot m^{-\alpha_3}~~{\rm for}~m_{b2} < m < m_{b3}\\
n(m) = A_4 \cdot m^{-\alpha_4}~~{\rm for}~m_{b3} < m < m_{b4}\\
n(m) = A_5 \cdot m^{-\alpha_5}~~{\rm for}~m > m_{b4},\\
\end{split}
\label{5s-eq_size_dist_m}
\end{equation} 
where $A_i$ are calculated from the corresponding continuity equations and from
\begin{equation}
\begin{split}
{m_{total} = \int\limits_{m_{min}}^{m_{max}}m\cdot n(m)dm = }\\
{{A_1\over-\alpha_1+2}(m_{b1}^{-\alpha_1+2}-m_{min}^{-\alpha_1+2})+}\\
{{A_2\over-\alpha_2+2}(m_{b2}^{-\alpha_2+2}-m_{b1}^{-\alpha_2+2})+}\\
{{A_3\over-\alpha_3+2}(m_{b3}^{-\alpha_3+2}-m_{b2}^{-\alpha_3+2})+}\\
{{A_4\over-\alpha_4+2}(m_{b4}^{-\alpha_4+2}-m_{b3}^{-\alpha_4+2})+}\\
{{A_5\over-\alpha_5+2}(m_{max}^{-\alpha_5+2}-m_{b4}^{-\alpha_5+2})},
\end{split}
\label{5s-A1}
\end{equation}
yielding,
\begin{equation}
\begin{split}
{A_1 = {m_{total} \over F}},
\end{split}
\label{5s-A1}
\end{equation}
with 
\begin{equation}
\begin{split}
{F = {m_{b1}^{-\alpha_1+2}-m_{min}^{-\alpha_1+2} \over -\alpha_1+2}+
m_{b1}^{-\alpha_1+\alpha_2} {m_{b2}^{-\alpha_2+2}-m_{b1}^{-\alpha_2+2} \over -\alpha_2+2}+}\\
m_{b1}^{-\alpha_1+\alpha_2} m_{b2}^{-\alpha_2+\alpha_3}{m_{b3}^{-\alpha_3+2}-m_{b2}^{-\alpha_3+2} \over -\alpha_3+2}+\\
m_{b1}^{-\alpha_1+\alpha_2} m_{b2}^{-\alpha_2+\alpha_3}m_{b3}^{-\alpha_3+\alpha_4}{m_{b4}^{-\alpha_4+2}-m_{b3}^{-\alpha_4+2} \over -\alpha_4+2}+\\
m_{b1}^{-\alpha_1+\alpha_2} m_{b2}^{-\alpha_2+\alpha_3}m_{b3}^{-\alpha_3+\alpha_4}m_{b4}^{-\alpha_4+\alpha_5}{m_{max}^{-\alpha_5+2}-m_{b4}^{-\alpha_5+2} \over -\alpha_5+2},
\end{split}
\label{5s-F}
\end{equation}
and 
\begin{equation}
\begin{split}
{A_2 = A_1\cdot m_{b1}^{-\alpha_1+\alpha_2}}\\
{A_3 = A_1\cdot m_{b1}^{-\alpha_1+\alpha_2}\cdot m_{b2}^{-\alpha_2+\alpha_3}}\\
{A_4 = A_1\cdot m_{b1}^{-\alpha_1+\alpha_2}\cdot m_{b2}^{-\alpha_2+\alpha_3}\cdot m_{b3}^{-\alpha_3+\alpha_4}}\\
{A_5 = A_1\cdot m_{b1}^{-\alpha_1+\alpha_2}\cdot m_{b2}^{-\alpha_2+\alpha_3}\cdot m_{b3}^{-\alpha_3+\alpha_4}\cdot m_{b4}^{-\alpha_4+\alpha_5}}.\\
\end{split}
\label{5s-A2A3A4A5}
\end{equation}

In terms if $r_{i}$, $A_{1}$ is given by
\begin{equation}
\begin{split}
{A_1 = {m_{total} \left({3\over4\pi\rho}\right)^{4-q_1\over 3} \over S},}
\end{split}
\label{5s-A_1mtot}
\end{equation} 

with
\begin{equation}
\begin{split}
{S = {3 \over 4-q_1}\left(r_{b1}^{4-q_1}-r_{min}^{4-q_1}\right)+r_{b1}^{q_2-q_1}{3 \over 4-q_2}\left(r_{b2}^{4-q_2}-r_{b1}^{4-q_2}\right)+}\\
{r_{b1}^{q_2-q_1}r_{b2}^{q_3-q_2}{3 \over 4-q_3}\left(r_{b3}^{4-q_3}-r_{b2}^{4-q_3}\right)+}\\
{r_{b1}^{q_2-q_1}r_{b2}^{q_3-q_2}r_{b3}^{q_4-q_3}{3 \over 4-q_4}\left(r_{b4}^{4-q_4}-r_{b3}^{4-q_4}\right)+}\\
{r_{b1}^{q_2-q_1}r_{b2}^{q_3-q_2}r_{b3}^{q_4-q_3}r_{b4}^{q_5-q_4}{3 \over 4-q_5}\left(r_{max}^{4-q_5}-r_{b4}^{4-q_5}\right)}.
\end{split}
\label{5s-A_1S}
\end{equation} 

The corresponding size distribution would be
\begin{equation}
\begin{split}
n(r) = A'_1 \cdot r^{-q_1}~~{\rm for}~r < r_{b1}\\
n(r) = A'_2 \cdot r^{-q_2}~~{\rm for}~r_{b1} < r < r_{b2}\\
n(r) = A'_3 \cdot r^{-q_3}~~{\rm for}~r_{b2} < r < r_{b3}\\
n(r) = A'_4 \cdot r^{-q_4}~~{\rm for}~r_{b3} < r < r_{b4}\\
n(r) = A'_5 \cdot r^{-q_5}~~{\rm for}~r > r_{b4},\\
\end{split}
\label{eq_size_dist_r}
\end{equation} 
where  $q_i = 3\alpha_i-2$ and $A'_i$ are derived from $n(m)dm=n(r)dr$ and given by
\begin{equation}
\begin{split}
{A'_i = 3^{\alpha_i}(4\pi\rho)^{-\alpha_i+1}A_i = 3^{{q_i+2}\over3}(4\pi\rho)^{{1-q_i}\over3}A_i}.
\end{split}
\label{5s-Aip}
\end{equation}
Substituting the expressions in Equations \eqref{5s-A1}, \eqref{5s-F}, and \eqref{5s-A2A3A4A5} into \eqref{5s-Aip}, and using $m_i = {4 \over 3}\pi\rho r_i^3$ and $\alpha_i = { q_i+2 \over 3}$, we get that the total number density of interstellar objects with radius $r \geqslant R$~($< r_{b1}$) is given by 
\begin{equation}
\begin{split}
{N_{\rm r \geqslant R} = \int\limits_{R}^{r_{b1}}A'_1r^{-q_1}dr+ \int\limits_{r_{b1}}^{r_{b2}}A'_2r^{-q_2}dr+}\\
{\int\limits_{r_{b2}}^{r_{rb3}}A'_3r^{-q_3}dr+ \int\limits_{r_{b3}}^{r_{rb4}}A'_4r^{-q_4}dr+ \int\limits_{r_{b4}}^{r_{max}}A'_5r^{-q_5}dr} =\\
{{A'_1\over-q_{1}+1}(r_{b1}^{-q_1+1}-R^{-q_1+1})+{A'_2\over-q_{2}+1}(r_{b2}^{-q_2+1}-r_{b1}^{-q_2+1})+}\\
{{A'_3\over-q_{3}+1}(r_{b3}^{-q_3+1}-r_{b2}^{-q_3+1})+{A'_4\over-q_{4}+1}(r_{b4}^{-q_4+1}-r_{b3}^{-q_4+1})+}\\
{{A'_5\over-q_{5}+1}(r_{max}^{-q_5+1}-r_{b4}^{-q_5+1})}.
\end{split}
\label{5s-N_r}
\end{equation}
\\
Dividing both terms of Equation \eqref{5s-N_r} by $m_{\rm total}$, we get
\begin{equation}
\begin{split}
{m_{total} = {N_{\rm r \geqslant R} \over T}}, \rm with \\
{T= {A''_1\over -q_1+1}(r_{b1}^{-q_1+1}-R^{-q_1+1})+}\\
{{A''_2\over -q_2+1}(r_{b2}^{-q_2+1}-r_{b1}^{-q_2+1})+}\\
{{A''_3\over -q_3+1}(r_{b3}^{-q_3+1}-r_{b2}^{-q_3+1})+}\\
{{A''_4\over -q_4+1}(r_{b4}^{-q_4+1}-r_{b3}^{-q_4+1})+}\\
{{A''_5\over -q_5+1}(r_{max}^{-q_5+1}-r_{b4}^{-q_5+1}),}
\end{split}
\label{5s-mtot}
\end{equation}
with 
\begin{equation}
\begin{split}
{A''_i = {A'_i \over m_{total}} = {A_i \over m_{total}} (4\pi\rho)^{1-q_i \over 3} 3^{q_i+2\over3}},
\end{split}
\label{5s-Appi}
\end{equation}
and $A_i$ given by \eqref{5s-A2A3A4A5}. Putting \eqref{5s-Appi} in terms of ${A_1 \over m_{total}}$ we get
\begin{equation}
\begin{split}
{A''_2 = {A_1 \over m_{total}} \left(4\pi\rho\right)^{1-q_1\over3}3^{2+q_1\over3}r_{b1}^{q_2-q_1}},\\
\end{split}
\label{5s-App2}
\end{equation}
\begin{equation}
\begin{split}
{A''_3 = {A_1 \over m_{total}} \left(4\pi\rho\right)^{1-q_1\over3}3^{2+q_1\over3}r_{b1}^{q_2-q_1}r_{b2}^{q_3-q_2}},\\
\end{split}
\label{5s-App3}
\end{equation}
\begin{equation}
\begin{split}
{A''_4 = {A_1 \over m_{total}} \left(4\pi\rho\right)^{1-q_1\over3}3^{2+q_1\over3}r_{b1}^{q_2-q_1}r_{b2}^{q_3-q_2}r_{b3}^{q_4-q_3}},\\
\end{split}
\label{5s-App4}
\end{equation}
\begin{equation}
\begin{split}
{A''_5 = {A_1 \over m_{total}} \left(4\pi\rho\right)^{1-q_1\over3}3^{2+q_1\over3}r_{b1}^{q_2-q_1}r_{b2}^{q_3-q_2}r_{b3}^{q_4-q_3}r_{b4}^{q_5-q_4}}.\\
\end{split}
\label{5s-App5}
\end{equation}
In Equations \eqref{5s-Appi}--\eqref{5s-App5}, ${A_1\over m_{total}}$ is given by Equations \eqref{5s-A_1mtot} and \eqref{5s-A_1S}. For a given $N_{\rm r \geqslant R}$ in pc$^{-3}$, and with $\rho$ in g cm$^{-3}$ and $R$ and $r_{i}$ in cm in all the equations above, the total mass density of solids $m_{\rm total}$ (g pc$^{-3}$) would be given by Equation \eqref{5s-mtot}.  

ORCID iDs
Amaya Moro-Martín https://orcid.org/0000-0001-
9504-8426


\begin{thebibliography}{99}
\bibitem[Andrews \& Williams(2007)]{2007ApJ...671.1800A} Andrews, S.~M., \& Williams, J.~P.\ 2007, ApJ, 671, 1800 
\bibitem[Bannister et al.(2017)]{2017ApJ...851L..38B} Bannister, M.~T., Schwamb, M.~E., Fraser, W.~C., et al.\ 2017, ApJL, 851, L38 
\bibitem[Bernstein et al.(2004)]{2004AJ....128.1364B} Bernstein, G.~M., Trilling, D.~E., Allen, R.~L., et al.\ 2004, AJ, 128, 1364 
\bibitem[Bland et al.(1996)]{1996MNRAS.283..551B} Bland, P.~A., Smith, T.~B., Jull, A.~J.~T., et al.\ 1996, MNRAS, 283, 551 
\bibitem[Bolin et al.(2018)]{2018ApJ...852L...2B} Bolin, B.~T., Weaver, H.~A., Fernandez, Y.~R., et al.\ 2018, ApJL, 852, L2 
\bibitem[Bottke et al.(2005)]{2005Icar..175..111B} Bottke, W.~F., Durda, D.~D., Nesvorn{\'y}, D., et al.\ 2005, Icar, 175, 111 
\bibitem[Brasser et al.(2006)]{2006Icar..184...59B} Brasser, R., Duncan, M.~J., \& Levison, H.~F.\ 2006, Icar, 184, 59 
\bibitem[{\'C}uk(2018)]{2018ApJ...852L..15C} {\'C}uk, M.\ 2018, ApJL, 852, L15 
\bibitem[Cumming et al.(2008)]{2008PASP..120..531C} Cumming, A., Butler, R. P.,  Marcy, G. W., Vogt, S. S., Wright, J. T. \ 2008, PASP, 120, 531
\bibitem[Do et al.(2018)]{2018ApJ...855L..10D} Do, A., Tucker, M.~A., \& Tonry, J.\ 2018, ApJL, 855, L10 
\bibitem[Drahus et al.(2018)]{2018NatAs...2..407D} Drahus, M., Guzik, P., Waniak, W., et al.\ 2018, NatAs, 2, 407 
\bibitem[Dressing \& Charbonneau(2015)]{2015ApJ...807...45D} Dressing, C.~D., \& Charbonneau, D.\ 2015, ApJ, 807, 45 
\bibitem[Dybczy{\'n}ski \& Kr{\'o}likowska(2018)]{2018A&A...610L..11D} Dybczy{\'n}ski, P.~A., \& Kr{\'o}likowska, M.\ 2018, A\&A, 610, L11 
\bibitem[Engelhardt et al.(2017)]{2017AJ....153..133E} Engelhardt, T., Jedicke, R., Vere{\v s}, P., et al.\ 2017, AJ, 153, 133 
\bibitem[Feng \& Jones(2018)]{2018ApJ...852L..27F} Feng, F., \& Jones, H.~R.~A.\ 2018, ApJL, 852, L27 
\bibitem[Fischer \& Valenti(2005)]{2005ApJ...622.1102F} Fischer, D.~A., \& Valenti, J.\ 2005, ApJ, 622, 1102 
\bibitem[Fitzsimmons et al.(2018)]{2018NatAs...2..133F} Fitzsimmons, A., Snodgrass, C., Rozitis, B., et al.\ 2018, NatAs, 2, 133 
\bibitem[Fraser \& Kavelaars(2009)]{2009AJ....137...72F} Fraser, W.~C., \& Kavelaars, J.~J.\ 2009, AJ, 137, 72 
\bibitem[Fraser et al.(2018)]{2018NatAs...2..383F} Fraser, W.~C., Pravec, P., Fitzsimmons, A., et al.\ 2018, NatAs, 2, 383 
\bibitem[Fuentes et al.(2009)]{2009ApJ...696...91F} Fuentes, C.~I., George, M.~R., \& Holman, M.~J.\ 2009, ApJ, 696, 91 
\bibitem[Gaidos(2018)]{2018MNRAS.477.5692G} Gaidos, E.\ 2018, MNRAS, 477, 5692 
\bibitem[Gaidos et al.(2017)]{2017RNAAS...1a..13G} Gaidos, E., Williams, J., \& Kraus, A.\ 2017, RNAAS, 1, 13 
\bibitem[Gomes et al.(2005)]{2005Natur.435..466G} Gomes, R., Levison, H.~F., Tsiganis, K., \& Morbidelli, A.\ 2005, Natur, 435, 466 
\bibitem[Hansen \& Zuckerman(2017)]{2017RNAAS...1a..55H} Hansen, B., \& Zuckerman, B.\ 2017, RNAAS, 1, 55 
\bibitem[Jackson et al.(2018)]{2018MNRAS.477L..85J} Jackson, A.~P., Tamayo, D., Hammond, N., Ali-Dib, M., \& Rein, H.\ 2018, MNRAS, 477, L85 
\bibitem[Jewitt et al.(2017)]{2017ApJ...850L..36J} Jewitt, D., Luu, J., Rajagopal, J., et al.\ 2017, ApJL, 850, L36 
\bibitem[Johansen \& Lambrechts(2017)]{2017AREPS..45..359J} Johansen, A., \& Lambrechts, M.\ 2017, AREPS, 45, 359 
\bibitem[Johnson et al.(2007a)]{2007ApJ...665..785J} Johnson, J.~A., Fischer, D.~A., Marcy, G.~W., et al.\ 2007, ApJ, 665, 785 
\bibitem[Johnson et al.(2007b)]{2007ApJ...670..833J} Johnson, J.~A., Butler, R.~P., Marcy, G.~W., et al.\ 2007, ApJ, 670, 833 
\bibitem[Kenyon et al.(2008)]{2008ssbn.book..293K} Kenyon, S.~J., Bromley, B.~C., O'Brien, D.~P., \& Davis, D.~R.\ 2008, The Solar System Beyond Neptune, M. A. Barucci, H. Boehnhardt, D. P. Cruikshank, and A. Morbidelli (eds.), University of Arizona Press, Tucson, pp.293-313 
\bibitem[Kroupa et al.(1993)]{1993MNRAS.262..545K} Kroupa, P., Tout, C.~A., \& Gilmore, G.\ 1993, MNRAS, 262, 545 
\bibitem[Lamy et al.(2004)]{2004come.book..223L} Lamy, P.~L., Toth, I., Fernandez, Y.~R., \& Weaver, H.~A.\ 2004, Comets II, M. C. Festou, H. U. Keller, and H. A. Weaver (eds.), University of Arizona Press, Tucson, pp. 223-264 
\bibitem[Laughlin \& Batygin(2017)]{2017RNAAS...1a..43L} Laughlin, G., \& Batygin, K.\ 2017, RNAAS, 1, 43 
\bibitem[Mamajek(2017)]{2017RNAAS...1a..21M} Mamajek, E.\ 2017, RNAAS, 1, 21 
\bibitem[Marcy et al.(2005)]{2005PThPS.158...24M} Marcy, G., Butler, R.~P., Fischer, D., et al.\ 2005, Progress of Theoretical Physics Supplement, 158, 24 
\bibitem[McGlynn \& Chapman(1989)]{1989ApJ...346L.105M} McGlynn, T.~A., \& Chapman, R.~D.\ 1989, ApJL, 346, L105 
\bibitem[Meech et al.(2017)]{2017Natur.552..378M} Meech, K.~J., Weryk, R., Micheli, M., et al.\ 2017, Natur, 552, 378 
\bibitem[Micheli et al.(2018)]{2018Natur.559..223M} Micheli, M., Farnocchia, D., Meech, K.~J., et al.\ 2018, Natur, 559, 223 
\bibitem[Morbidelli et al.(2005)]{2005Natur.435..462M} Morbidelli, A., Levison, H.~F., Tsiganis, K., \& Gomes, R.\ 2005, Natur, 435, 462 
\bibitem[Moro-Mart{\'{\i}}n et al.(2007)]{2007ApJ...658.1312M} Moro-Mart{\'{\i}}n, A., Carpenter, J.~M., Meyer, M.~R., et al.\ 2007, ApJ, 658, 1312 
\bibitem[Moro-Mart{\'{\i}}n et al.(2009)]{2009ApJ...704..733M} Moro-Mart{\'{\i}}n, A., Turner, E.~L., \& Loeb, A.\ 2009, ApJ, 704, 733 
\bibitem[Moro-Mart{\'{\i}}n(2013)]{2013pss3.book..431M} Moro-Mart{\'{\i}}n, A.\ 2013, Planets, Stars and Stellar Systems.~Volume 3: Solar and Stellar Planetary Systems, T. D. Oswalt, L. M. French, P. Kalas (eds.), Springer Science+Business Media, Dordrecht, pp. 431-487
\bibitem[Moro-Mart{\'{\i}}n et al.(2015)]{2015ApJ...801..143M} Moro-Mart{\'{\i}}n, A., Marshall, J.~P., Kennedy, G., et al.\ 2015, ApJ, 801, 143 
\bibitem[O'Brien et al.(2007)]{2007Icar..191..434O} O'Brien, D.~P., Morbidelli, A., \& Bottke, W.~F.\ 2007, Icar, 191, 434 
\bibitem[Plane et al.(2018)]{2018SSRv..214...23P} Plane, J.~M.~C., Flynn, G.~J., M{\"a}{\"a}tt{\"a}nen, A., et al.\ 2018, SSRv, 214, \#23 
\bibitem[Rafikov(2018)]{2018ApJ...861...35R} Rafikov, R.~R.\ 2018, ApJ, 861, 35 
\bibitem[Raymond et al.(2018a)]{2018ApJ...856L...7R} Raymond, S.~N., Armitage, P.~J., \& Veras, D.\ 2018, ApJL, 856, L7 
\bibitem[Raymond et al.(2018b)]{2018MNRAS.476.3031R} Raymond, S.~N., Armitage, P.~J., Veras, D., Quintana, E.~V., \& Barclay, T.\ 2018, MNRAS, 476, 3031 
\bibitem[Scalo(1986)]{1986FCPh...11....1S} Scalo, J.~M.\ 1986, FCPh, 11, 1 
\bibitem[Schlichting et al.(2013)]{2013AJ....146...36S} Schlichting, H.~E., Fuentes, C.~I., \& Trilling, D.~E.\ 2013, AJ, 146, 36 
\bibitem[Trilling et al.(2017)]{2017ApJ...850L..38T} Trilling, D.~E., Robinson, T., Roegge, A., et al.\ 2017, ApJL, 850, L38 
\bibitem[Tsiganis et al.(2005)]{2005Natur.435..459T} Tsiganis, K., Gomes, R., Morbidelli, A., \& Levison, H.~F.\ 2005, Natur, 435, 459 
\bibitem[Williams(2017)]{Williams2017}Williams, G., Minor Planet Electronic Circular 2017-U181 (October 25)
\bibitem[Winn \& Fabrycky(2015)]{2015ARA&A..53..409W} Winn, J.~N., \& Fabrycky, D.~C.\ 2015, ARA\&A, 53, 409 
\bibitem[Wyatt et al.(2017)]{2017MNRAS.464.3385W} Wyatt, M.~C., Bonsor, A., Jackson, A.~P., Marino, S., \& Shannon, A.\ 2017, MNRAS, 464, 3385 
\bibitem[Zhang(2018)]{2018ApJ...852L..13Z} Zhang, Q.\ 2018, ApJL, 852, L13 
\bibitem[Portegies Zwart et al.(2018)]{2018MNRAS.479L..17P} Portegies Zwart, S., Torres, S., Pelupessy, I., B{\'e}dorf, J., \& Cai, M.~X.\ 2018, MNRAS, 479, L17 

\end{thebibliography}
\end{document}